\documentclass{aa}  

\usepackage{natbib}
\usepackage{mathtools}
\usepackage{array}
\usepackage{caption}
\usepackage{subcaption}
\usepackage{wrapfig}
\usepackage{xcolor}
\usepackage{ulem}
\usepackage{graphicx}
\usepackage{txfonts}
\usepackage[citecolor=blue]{hyperref}
\usepackage{multirow}

\begin{document}

   \title{Ly-alpha emission reveals two satellite halos around massive groups at $z\sim$ 3: The puzzling case of a quiescent central galaxy}
   \titlerunning{Ly-alpha emission reveals two satellite halos around massive groups at $z\sim$ 3}

   \author{Sicen Guo\inst{1}
          \and
          Emanuele Daddi\inst{1}
          \and
          Raphael Gobat\inst{2}
          \and
          Nikolaj B. Sillassen\inst{3,4}
          \and
          Chiara D’Eugenio\inst{1}
          \and
          R. Michael Rich\inst{5}
          \and
          Guillaume Elias\inst{1}
          \and
          Manuel Aravena\inst{6,7}
          \and
          Franziska Bruckmann\inst{1}
          \and
          Camila Correa\inst{8}
          \and
          Ivan Delvecchio\inst{9}
          \and
          David Elbaz\inst{1}
          \and
          Sofia G. Gallego\inst{10}
          \and
          Fabrizio Gentile\inst{1,11}
          \and
          Shuowen Jin\inst{3,4}
          \and
          Boris S. Kalita\inst{12,13}
          \and
          James D. Neill\inst{14}
          \and
          Manuel Solimano\inst{6}
          \and
          Francesco Valentino\inst{4,15}
          \and
          Tao Wang\inst{16,17}
          }

   \institute{CEA, IRFU, DAp, AIM, Université Paris-Saclay, Université de Paris, Sorbonne Paris Cité, CNRS, F-91191 Gif-sur-Yvette, France\\
              \email{sicen.guo@cea.fr}
              \and
              Instituto de Física, Pontificia Universidad Católica de Valparaíso, Casilla 4059, Valparaíso, Chile
              \and
              Cosmic Dawn Center (DAWN), Denmark
              \and
              DTU Space, Technical University of Denmark, Elektrovej 327, DK-2800 Kgs. Lyngby, Denmark
              \and
              Department of Physics and Astronomy, UCLA PAB 430 Portola Plaza Los Angeles, CA 90095-1547
              \and
              Instituto de Estudios Astrof\'{\i}cos, Facultad de Ingenier\'{\i}a y Ciencias, Universidad Diego Portales, Av. Ej\'ercito 441, Santiago, Chile
              \and
              Millenium Nucleus for Galaxies (MINGAL)
              \and
              Leiden Observatory, Leiden University, PO Box 9513, 2300 RA Leiden, the Netherlands
              \and
              INAF – Osservatorio di Astrofisica e Scienza dello Spazio di Bologna, Via Gobetti 93/3, I-40129 Bologna, Italy
              \and
              Université Paris Cité, CNRS(/IN2P3), Astroparticule et Cosmologie, F-75013 Paris, France
              \and
              INAF- Osservatorio di Astrofisica e Scienza dello Spazio, Via Gobetti 93/3, I-40129, Bologna, Italy
              \and
              Kavli Institute for Astronomy and Astrophysics, Peking University, Beijing 100871, People{\textquotesingle}s Republic of China
              \and
              Kavli Institute for the Physics and Mathematics of the Universe, The University of Tokyo, Kashiwa, 277-8583, Japan
              \and
              California Institute of Technology, 1200 East California Boulevard, MC 278-17, Pasadena, CA 91125, USA
              \and
              European Southern Observatory, Karl-Schwarzschild-Str. 2, D85748 Garching bei Munchen, Germany
              \and
              School of Astronomy and Space Science, Nanjing University, Nanjing 210093, China
              \and
              Key Laboratory of Modern Astronomy and Astrophysics (Nanjing University), Ministry of Education, Nanjing 210093, China
             }

   \date{Received ***; accepted ***}

  \abstract
{We present the discovery and characterisation of two Ly$\alpha$ nebulae (LANs), RO-1001-Sat and RO-0959-Sat, as satellite structures of two giant LANs at $z=2.920$ and 3.092. These two satellite LANs are found neighbouring two out of four known giant LANs at $z\sim3$ in our Multi-Unit Spectroscopic Explorer (MUSE) follow-up observations, reinforcing the idea that Ly$\alpha$ emission can be used as a tracer of massive dark matter halos at high $z$. This high occurrence of massive satellite halos is in agreement with simulations. With sizes of $\simeq80\times160$ and $80\times100~\mathrm{pkpc}^2$, the two nebulae are both $\sim$ 300 pkpc from the main LANs. With Ly$\alpha$ emission only shifted by $\simeq100-300$ km s$^{-1}$ between each of the two pairs, these two satellite structures are likely connected to their main halos by the large-scale structure. RO-1001-Sat and RO-0959-Sat are estimated to have log$(M_\mathrm{h}/M_\odot)\simeq13.2\pm0.3$ and $12.8\pm0.3$, putting them potentially close to the regime of cold-mode accretion according to several models, which suggests that cold streams should be able to penetrate the virial radii to reach the halo centres without being shock-heated. The central brightest galaxies in the two halos are morphologically distinct despite having a similar stellar mass of $\sim10^{11}M_\odot$, one being an elliptical quiescent galaxy in RO-1001-Sat and the other being a dusty star-forming spiral in RO-0959-Sat. Intriguingly, the quiescent galaxy aligns well with the peak of the LAN as well as the potential well of the host halo, making it the first clear-cut case in which the cold gas ought to be accreting onto the galaxy but with no observable star formation, either due to morphological quenching or, more likely, radio-mode feedback from an active galactic nucleus, as supported by excess yet weak radio emission. Finally, we show a tentative detection of a Ly$\alpha$ filament connecting RO-1001 and RO-1001-Sat. This work shows how panoramic MUSE (and in the future, BlueMUSE) observations of massive halo seeds can be used to efficiently search for additional halos, unveiling their large-scale structure and enabling the study of Ly$\alpha$-selected galaxy groups.
}


   \maketitle

\section{Introduction} \label{sec:introduction}

Galaxy clusters reside in overdense dark matter halos connected by cosmic filaments that form the large-scale cosmic web as predicted by cold dark matter cosmology (\citealt{Peebles82, Blumenthal84, Davis85}). Many cases of cluster pairs connected by such filaments are known in the local Universe (e.g. \citealt{Planck13, Bonjean18, Govoni19, Tanimura20, Isopi25}). Since galaxy clusters collapse from the densest regions of the total mass field, they are a biased tracer of the underlying matter distribution (\citealt{Kaiser84, Bardeen86, Cole89, Sheth99, Desjacques18} for a review). This bias is predicted to be a strong function of redshift, becoming stronger at higher redshifts and easing over time towards unity, since the first structures ought to be formed at the highest density peaks of the matter distribution (\citealt{Fry96, Tegmark98, Mo02}). Thus, for a given halo mass, $M_\mathrm{h}$, we would naturally expect to see a higher clustering of structures in the distant Universe. In addition, halo abundances are expected to undergo rapid growth down to $z\sim3$ for halos of $M_\mathrm{h} \gtrsim 10^{13}M_\odot$ \citep{Mo02}, implying more frequent mass assembly and mergers of substructures. This again suggests that more protocluster pairs (i.e. the counterpart of the above-mentioned pairs of local clusters) should exist in the early Universe. However, despite the increased search for protoclusters in the past decade (e.g. \citealt{Toshikawa16, Toshikawa18, Higuchi19, Wang21, Ramakrishnan24}), there is still a lack of direct detections of cluster and group pairs at higher $z$. One of the reasons may be that the typical $M_\mathrm{h}$ at high $z$ is intrinsically lower, since most of the groups are in their early stage of formation, and thus the bias of what we can observe does not necessarily increase with $z$. Alternatively, pairs of high-$z$ massive halos might already exist in previous observations, but were not identified owing to a lack of $M_\mathrm{h}$ associations with these structures (such as what has been done in \citealt{Wang16, Daddi21, Daddi22a, Sillassen22, Mei23, Sillassen24}). Another way to trace the large-scale structure around high-z overdensities is to look for giant ($\gtrsim$100 kpc) Ly$\alpha$ nebulae (LANs), which are found to be helpful in tracing protoclusters (e.g. \citealt{Valentino16, Daddi21, Daddi22a}).

Cosmic filaments between protoclusters are even more challenging to detect due to the much lower surface brightness (SB). Nevertheless, many filamentary structures have been reported at $z>2$ in the intergalactic medium (IGM) scales. Such observations were traditionally done on galaxy-quasar (QSO) systems through absorption lines in the QSO spectra (see \citealt{Rauch98} \& \citealt{Meiksin09} for reviews) where valuable information was lost by integration along the line of sight (LOS). With spectrographs such as Keck Cosmic Web Imager (KCWI) and Multi-Unit Spectroscopic Explorer (MUSE) employed, direct mapping of at least the brightest components of the cosmic web has been made possible primarily through fluorescent Ly$\alpha$ emission around or between QSOs (e.g. \citealt{Cantalupo14, Battaia19, Tornotti25a}) as well as in and around overdense environments (e.g. \citealt{Bacon21, Banerjee25}). Multiple giant LANs spanning over 100 kpc that are likely powered by either active galactic nucleus (AGN) photoionisation or collisional excitation through cosmological cold flow accretion have been discovered at $z\approx2-3$ permeating the intergalactic space inside protoclusters (\citealt{Valentino16, Daddi21, Daddi22a}). Megaparsec-scale segments of the cosmic web in Ly$\alpha$ emission have also been reported by \cite{Umehata19} and \cite{Tornotti25b}. However, the detection of Ly$\alpha$ filaments connecting massive non-QSO hosting galaxy groups or clusters at $z\gtrsim2$ still remains elusive. This is expected, since such cosmic filaments have extremely low SB(Ly$\alpha$) of $10^{-20}-10^{-19}\rm{erg}~s^{-1}\rm{cm}^{-2}\rm{arcsec}^{-2}$ as predicted by simulations (e.g. \citealt{Gould96, Liu25}) at $z\sim3$ and beyond, well below the detection limit of current instrumentation. In comparison, \citet{Gallego18} reach a 2$\sigma$ SB level of $\approx0.44\times10^{-20}\rm{erg}~s^{-1}\rm{cm}^{-2}\rm{arcsec}^{-2}$ for Ly$\alpha$ filaments at $3<z<4$ after stacking 390 sub-cubes from the deepest MUSE observations oriented by the positions of Ly$\alpha$ emitter (LAE) neighbours. Both the MUSE eXtremely Deep Field (MXDF; \citealt{Bacon21, Bacon23}) and the MUSE Ultra Deep Field (MUDF; \citealt{Lusso19, Fossati19}), consisting of $\approx$140-h total exposure each, have also reached comparable sensitivities. Searching for fainter and more extended segments of the cosmic filaments is crucial to study the large-scale structure, as half of the matter in the Universe is predicted to be distributed along the filaments by simulations (e.g. \citealt{Cautun14}). Moreover, it is important for our understanding of galaxy evolution, since it has been shown that there are significant correlations between galaxies and the large-scale structure they reside in (e.g. \citealt{Balogh04, Bamford09, Blanton09, Peng10}).

Furthermore, cold gas is predicted to be transported into dark matter halos through streams in filaments according to the cold accretion model (e.g. \citealt{Birnboim03, Kere05, Dekel06, Dekel09, Dekel13}) so that galaxies residing in the halos can sustain star formation much longer than the typical gas consumption timescales (e.g. \citealt{Lilly13, Walter20}). Based on this cold accretion model, the baryonic accretion rate (BAR) of cold gas that can be used to form stars onto dark matter halos should increase with both the halo mass and redshift up to $M_\mathrm{h}\simeq M_{\rm stream}$, above which the cold accretion decreases (e.g. \citealt{Dekel06, Dekel09, Daddi22a}). If this theory holds, the cold accretion rates are higher in high-$z$ massive clusters where a larger amount of cold gas that is ready to form stars is expected to be present. Given that at high $z$ the environmental quenching may not have an impact on galaxy evolution as significant as in the local Universe, findings of quiescent galaxies (QGs) in such overdensities in the distant Universe (e.g. \citealt{Kalita21, Kubo21, McConachie22, Kakimoto24, Ito25}) highlight the necessity of alternative quenching mechanisms, such as AGN feedback (e.g. \citealt{Sanders88, Matteo05, Croton06, Hopkins06, Fabian12}) and morphological quenching (e.g. \citealt{Martig09, Genzel14, Tacchella15, Gobat18}). More intriguingly, some QGs have even been found in regions suspected to host large cold gas reservoirs, traced by Ly$\alpha$ ($\sim10^4$ K) or CO ($\lesssim10^2$ K) emission. \citet{Kubo21} reported a massive QG in a protocluster at $z\sim3$ at the edge of a patchy LAN \citep{Umehata19}. Recent follow-up ALMA observations detect a molecular gas reservoir around this galaxy \citep{Umehata25}. Another similar case is reported by \citet{Kalita21}, in which a massive QG is discovered within 10$\arcsec$ from the peak of the giant LAN in the galaxy group RO-1001, also at $z\sim3$. On the other hand, there are cases of extremely star-forming protoclusters such as SPT2349-56, which hosts a massive molecular gas reservoir \citep{Zhou25} but is associated with relatively faint Ly$\alpha$ emission \citep{Apostolovski24}.

In this paper, we present the discovery of two extended satellite Ly$\alpha$ emissions, RO-1001-Sat and RO-0959-Sat, at $\sim$1 cMpc from two main LANs, RO-1001 and RO-0959. We also report the finding of another QG in a LAN as described above, but in this case more peculiarly sitting right at the centre of RO-1001-Sat. These four LANs trace four protoclusters at $z\sim3$ that form two above-mentioned overdensity pairs at high $z$. RO-1001 is reported by \citet{Daddi21} to host three Ly$\alpha$ filaments converging on its potential well. The Ly$\alpha$ emissions in both RO-1001 and RO-0959 have been previously observed by KCWI. Thanks to the larger field of view (FOV) of MUSE, we are able to identify these two satellite structures.

\begingroup
\renewcommand{\arraystretch}{1.4}
\begin{table*}[t]
    \caption{\label{tab:obs_details}Two satellite LANs studied in this work and their main nebulae.}
    \centering
    \begin{tabular}{lcccccccc}
        \hline
        ID & R.A. & Dec. & $z$\tablefootmark{a} & log$(M_\mathrm{h})$\tablefootmark{b} & $R_\mathrm{vir}$\tablefootmark{c} & log($L_{\mathrm{Ly}\alpha}$) & $d_\mathrm{proj}$\tablefootmark{d} & $T_\mathrm{int}$ \\
        & & & & $(M_\odot)$ & (kpc) & (erg s$^{-1}$) & (pkpc) & (h) \\ \hline
        RO-1001 & 10:01:23.06 & 02:20:04.9 & 2.916 & 13.6 $\pm$ 0.2 & 269 $\pm$ 41 & 44.0 & - & 7.9 \\
        RO-1001-Sat & 10:01:24.25 & 02:20:38.0 & 2.920 & 13.2 $\pm$ 0.3 & 197 $\pm$ 45 & 43.4 & 311 & 7.9 \\ \hline
        RO-0959 & 09:59:59.48 & 02:34:41.7 & 3.091 & 12.8 $\pm$ 0.3 & 139 $\pm$ 32 & 44.0 & - & 6.3 \\
        RO-0959-Sat & 09:59:57.04 & 02:34:36.9 & 3.092 & 12.8 $\pm$ 0.3 & 139 $\pm$ 32 & 42.9 & 276 & 6.3 \\ \hline
    \end{tabular}
    \tablefoot{\tablefoottext{a}{$z$ was calculated from the flux-weighted peak in Ly$\alpha$ spectra integrated across each nebula.} \tablefoottext{b}{$M_\mathrm{h}$ (and also $L_\mathrm{Ly\alpha}$) from \cite{Daddi22a} was used for the main nebulae, while the $M_\mathrm{h}$ of the satellites was estimated using the methods in \cite{Sillassen24} (see Sect. \ref{sec:Mh}).} \tablefoottext{c}{$R_\mathrm{vir}$ was derived from $M_\mathrm{h}$ using the $M_\mathrm{h}-R_\mathrm{vir}$ relation in \citet{Goerdt10}.} \tablefoottext{d}{$d_\mathrm{proj}$ was measured between the two Ly$\alpha$ peaks in each pair of LANs.}}
\end{table*}
\endgroup

This paper is organised as follows. In Sect. \ref{sec:data_redection} we describe the MUSE data reduction and noise calibration procedures we employed. In Sect. \ref{sec:results} we present the estimates of halo masses of the two new satellite LANs and near neighbour probabilities in simulations. We show the characterisation of the central brightest galaxies in the two LANs and search for filaments between the two halo pairs. In Sect. \ref{sec:discussion}, we explore the possible quenching mechanisms of the central QG in RO-1001-sat and the powering sources of the two new satellite LANs. We also discuss the implications of the discovery of these two LANs and the tentative filament detection. We summarise and conclude in Sect. \ref{sec:conclusions}. We adopt a standard $\Lambda$CDM cosmology with $H_0=70~\mathrm{km~s}^{-1}\mathrm{Mpc}^{-1},~\Omega_\mathrm{m}=0.3,~\Omega_\Lambda=0.7$, and a \cite{Chabrier03} initial mass function. 

\section{MUSE data reduction} \label{sec:data_redection}
RO-1001 and RO-0959 are part of a MUSE follow-up (PI: R. Gobat, proposal ID:110.23ST) of five galaxy groups at $z\sim$ 2-3.5. We observed Ly$\alpha$ lines in four of the structures, while the fifth does not have a redshift high enough for Ly$\alpha$ to be observable. The observations were performed using the Wide Field Mode with adaptive optics (WFM-AO-E) covering a $1\arcmin\times1\arcmin$ FOV. RO-1001 was observed between June 9 2021 and January 18 2023 with a total exposure time of 7.9h and seeing in the range of 0.3-2.8\arcsec. RO-0959 was observed between December 28 2022 and January 25 2023 with a total exposure time of 6.3h and seeing of 0.26-3.2\arcsec. The properties of these two pairs of LANs are listed in Table \ref{tab:obs_details}.

We reduced the raw datasets using the \texttt{esoreflex} pipeline for MUSE from ESO (v2.11.5, \citealt{Weilbacher20}) by passing to the pipeline a galaxy mask from the COSMOS2020 Classic catalogue (COSMOS2020 hereafter, \citealt{Weaver22}) and setting the \texttt{autocalib} parameter to \texttt{deepfield} for the \texttt{muse\_scipost} recipe. The output cubes from the pipeline contain both data and variance cubes. We noticed some stripe patterns in the white image (i.e. an image obtained by collapsing the wavelength axis of the data cube) of the combined cube from the pipeline, as well as the fluctuating zero-levels in the spectra extracted especially from the longer wavelengths. This was introduced by the differences between sensitivities in different detectors that were not resolved by flat-fielding of the pipeline. To deal with this issue, we created a super-sky cube \citep{Bacon23} by combining sky backgrounds in frames that have 30 min exposure time from all five fields in our sample, totalling 33.5h. This super-sky cube was then subtracted from all the individual frames to improve the flat-field correction before subsequent noise calibrations and final combinations.

We calibrated the variance cubes after masking out spaxels on all objects and LANs, the shapes of which were drawn from narrow-band images of the previous KCWI observations, using 3D masks obtained by extending the 2D spatial masks across all spectral channels. The noise calibrations were done at each wavelength (i.e. per layer) and then at each location (i.e. per spaxel) by scaling the variance to re-normalise the distribution of the background pixels in the signal-to-noise (S/N) space to be a standard Gaussian. We then used ZAP (the Zurich Atmospheric Purge; \citealt{Soto16}) to enhance the sky subtraction by the pipeline. For the combined cube in each field, we performed pixel-by-pixel 3$\sigma$-clippings of all frames weighted by their calibrated variance cubes and corrected the zero sky level with its median value at each wavelength if a positive background level was present. Finally, we used \texttt{CWITools} (\citealt{O'Sullivan20, Martin19}) to produce a new narrow-band image and a corresponding 3D mask of the LAN from each combined cube and iterated the above noise calibration procedures with this new 3D Ly$\alpha$ mask in each field.

After the iteration, we created a continuum background model using \texttt{CWITools} by median filtering with a window width of 31$\AA$ in each field and subtracted it from the final combined data cube to remove any continuum sources. As no bright UV source is present in either of the two fields, no obvious continuum residuals were seen in the continuum-subtracted cubes, and no further masking was needed. To search for extended Ly$\alpha$ emission, we applied adaptive kernel smoothing (AKS, \citealt{Martin19}), in which Gaussian kernels are used to smooth the data cube both spatially and spectrally. We started with a kernel size of 0.8$''$ (2 $\AA$) that increased in each iteration until it reached the maximum spatial, $\sim11''$, (spectral, 30$\AA$) smoothing scale. In each iteration, smoothed voxels above 2.5$\sigma$ were extracted to the AKS data cube. As a result, brighter areas were smoothed less and fainter areas more. Then we constructed a segmentation cube where the voxels belonging to the same extended emission were identified to produce the Ly$\alpha$ SB, first and second moment maps of each LAN. The S/N maps of the AKS images of the two LANs are shown in Appendix \ref{app:snr_maps}. Using the SB map as a mask for each nebula, we extracted the Ly$\alpha$ and its noise spectra from the continuum-subtracted data cube and the calibrated variance cube, respectively. We re-normalised the noise using the same approach as described above by rescaling the background level for it to have a standard Gaussian distribution in the S/N space. Further details on the data reduction and analysis will be published in a forthcoming paper (Guo et al., in prep).

\section{Results} \label{sec:results}
\subsection{Satellite halos} \label{sec:satellites}
\begin{figure*}
    \sidecaption
    \includegraphics[width=0.7\linewidth]{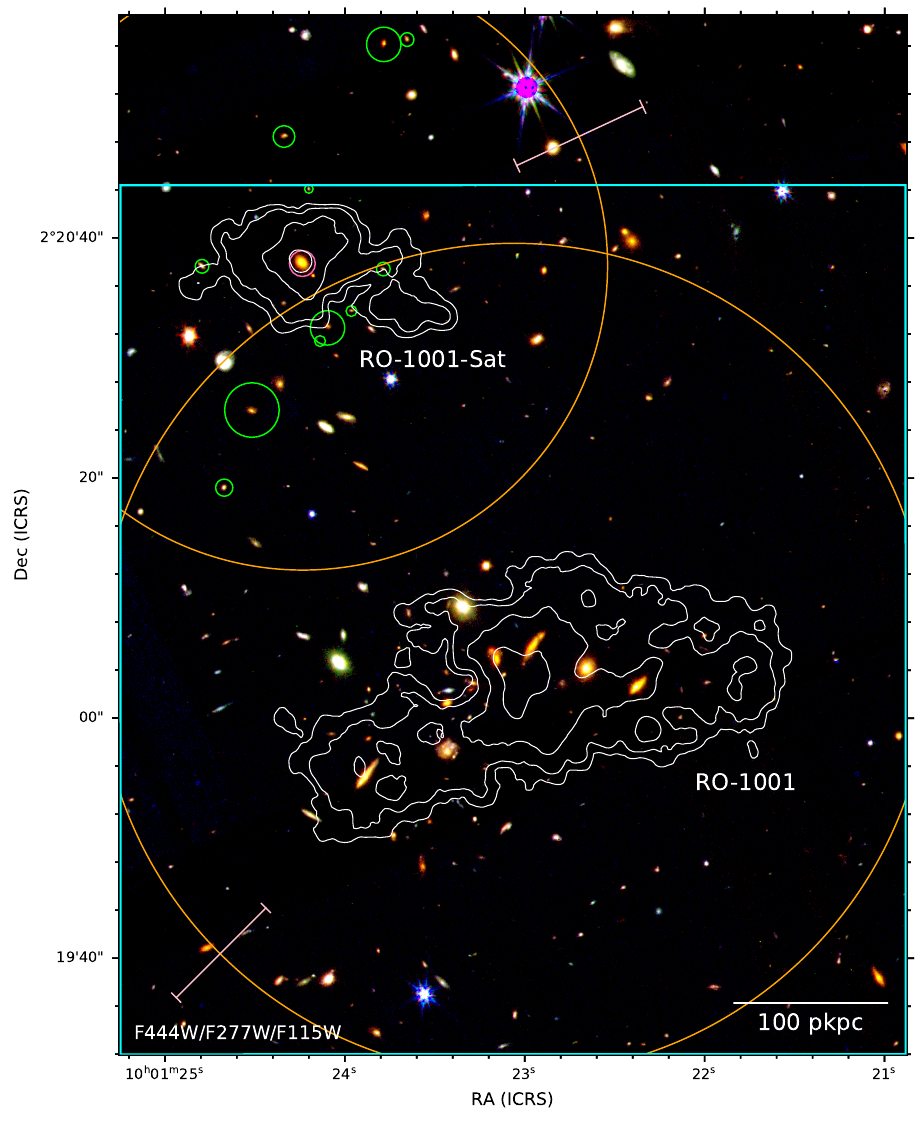}
    \caption{Colour image of RO-1001 and RO-1001-Sat from F444W, F277W and F115W bands of JWST NIRCam. The cyan box shows the $1.1\arcmin\times 1.2\arcmin$ FOV of the MUSE mosaic. The orange circles show the virial radii of the two halos with the errors illustrated by the pink segment on each circle, being $197\pm45$ and $269\pm41$ kpc for RO-1001-Sat and RO-1001. The contours show the Ly$\alpha$ SB distributions in both nebulae displayed in steps of log$\rm (SB_{Ly\alpha}/\mathrm{erg~s}^{-1}\mathrm{cm}^{-2}\mathrm{arcsec}^{-2})$ = (-18.5, -18.0, -17.5, -17.0). The lowest contour level corresponds to S/N=5. Candidate member galaxies of the satellite halo in Table \ref{tab:galaxies_ro1001} are marked by green circles. The central brightest galaxy (log$(M_*/M_\odot)\simeq11.3$) is highlighted by the pink circle. The radii of the green circles are proportional to the stellar masses of the galaxies, ranging from log$(M_*/M_\odot)\simeq9.5-10.3$.}
    \label{fig:rgb_ro1001}
\end{figure*}

\begin{figure*}
    \sidecaption
    \includegraphics[width=0.7\linewidth]{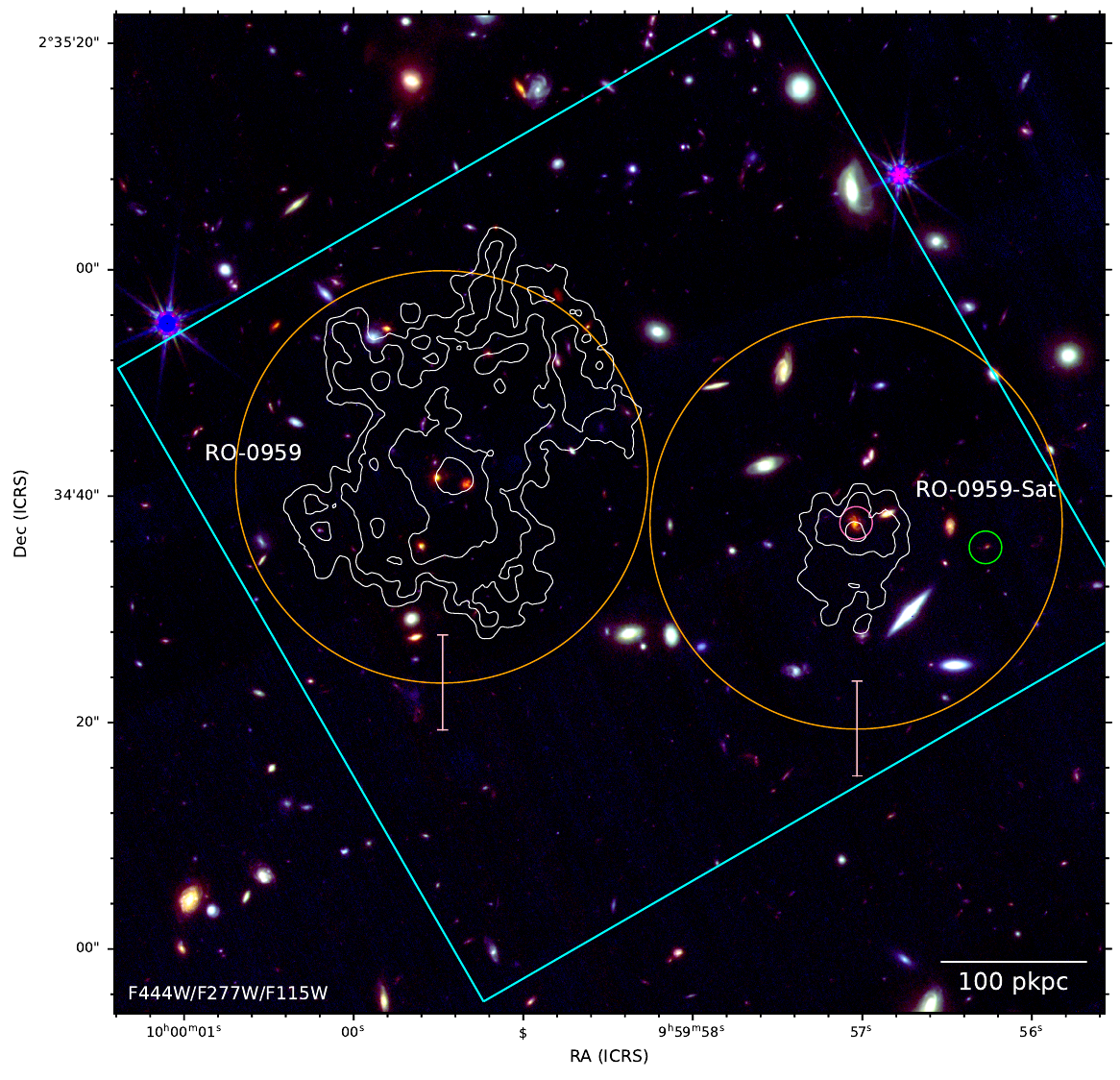}
    \caption{Colour image of RO-0959 and RO-0959-Sat from F444W, F277W and F115W bands of JWST NIRCam. The cyan box shows the $1.1\arcmin\times 1.1\arcmin$ FOV of the MUSE mosaic. The orange circles show the virial radii of the two halos with the errors illustrated by the pink segment on each circle, being $139\pm32$ kpc for both RO-0959-Sat and RO-0959. The contours show the Ly$\alpha$ SB distributions in both nebulae displayed in steps of log$\rm (SB_{Ly\alpha}/\mathrm{erg~s}^{-1}\mathrm{cm}^{-2}arcsec^{-2})$ = (-18.5, -18.0, -17.5, -17.0). The lowest contour level corresponds to S/N=3. The central and the second candidate member galaxies in Table \ref{tab:galaxies_RO0959} with log$(M_*/M_\odot)\simeq11.2$ and 9.9 are highlighted by the pink and green circles.}
    \label{fig:rgb_ro0959}
\end{figure*}

\begin{figure*}
    \centering
    \begin{subfigure}{0.48\linewidth}
        \centering
        \includegraphics[width=\linewidth]{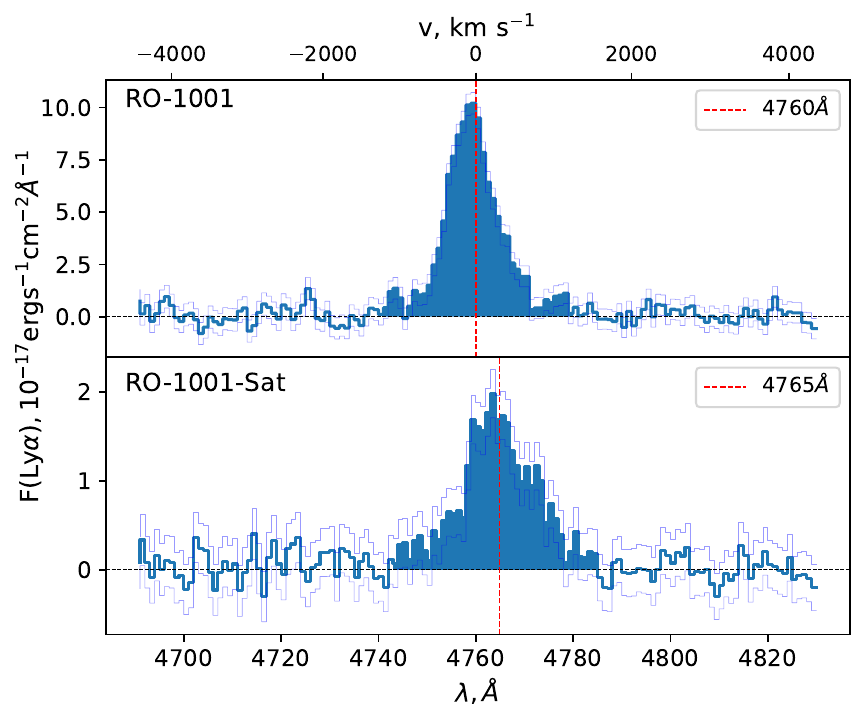}
        \label{fig:lya_spec_ro1001}
    \end{subfigure}
    \begin{subfigure}{0.48\linewidth}
        \centering
        \includegraphics[width=\linewidth]{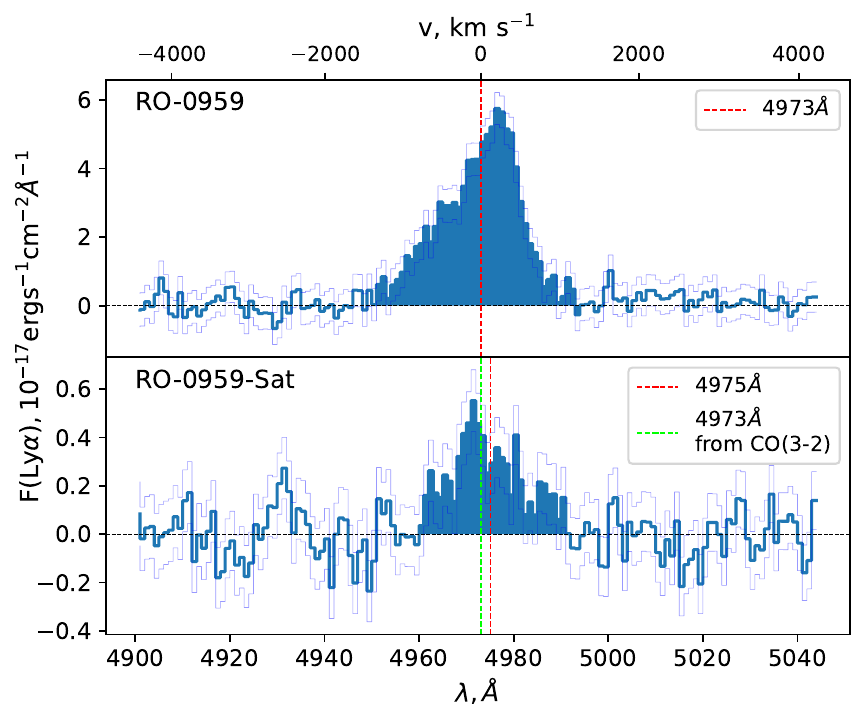}
        \label{fig:lya_spec_ro0959}
    \end{subfigure}
    \caption{Ly$\alpha$ spectra integrated from the main and satellite nebulae of RO-1001 (\textit{left}) and RO-0959 (\textit{right}) after noise re-normalisation. Dashed red lines mark the flux-weighted peak of each spectrum. The dashed green line in the lower right panel shows the expected position of the Ly$\alpha$ peak from the CO(3-2) emission in RO-0959-Sat.}
    \label{fig:lya_specs}
\end{figure*}

\begin{figure}
    \centering
    \begin{subfigure}{0.5\textwidth}
        \centering
        \includegraphics[width=\textwidth]{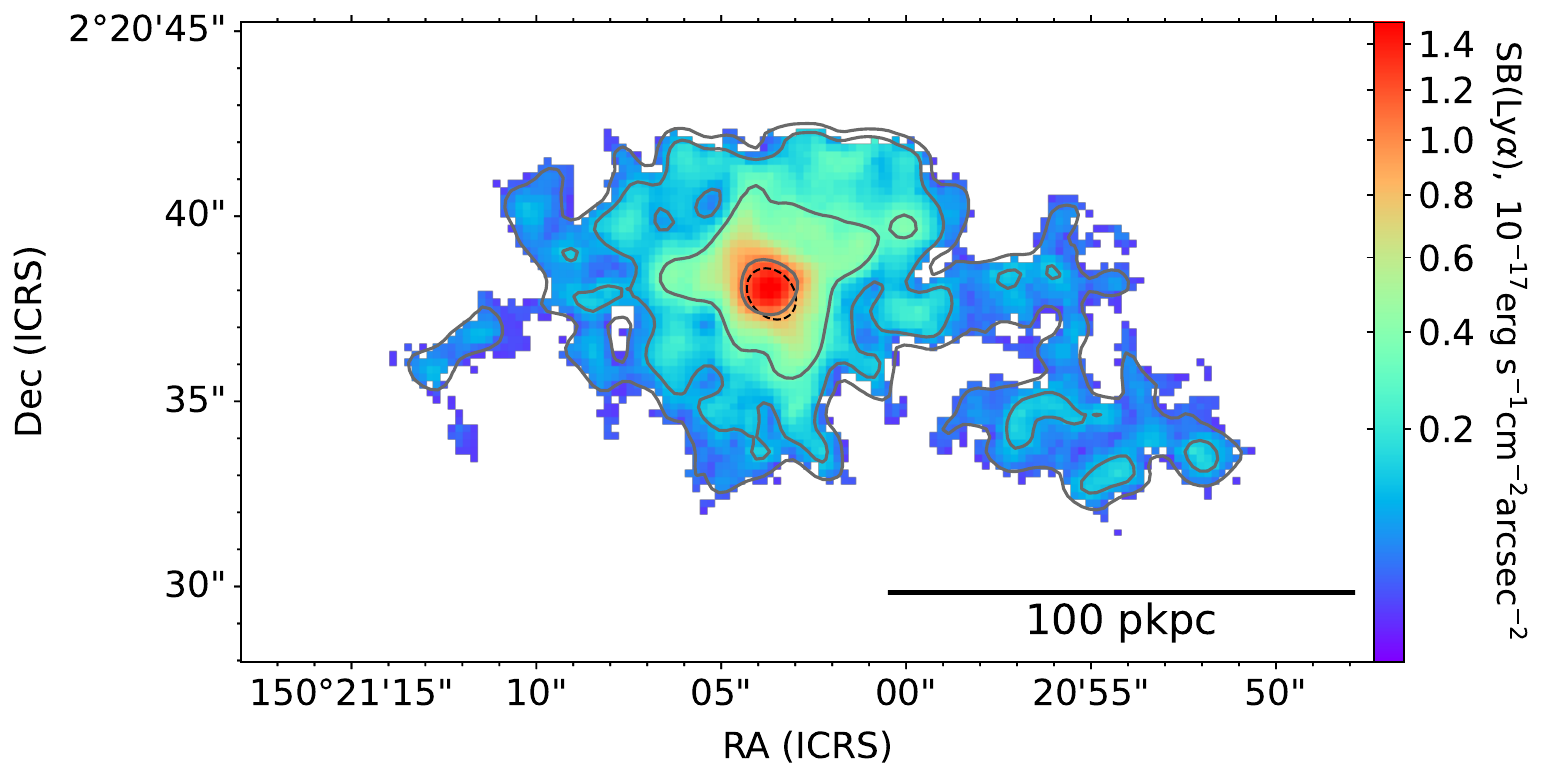}
        \label{fig:sb-ro1001sat}
        \vspace{-1.125cm}
    \end{subfigure}
    \begin{subfigure}{0.5\textwidth}
        \centering
        \includegraphics[width=\textwidth]{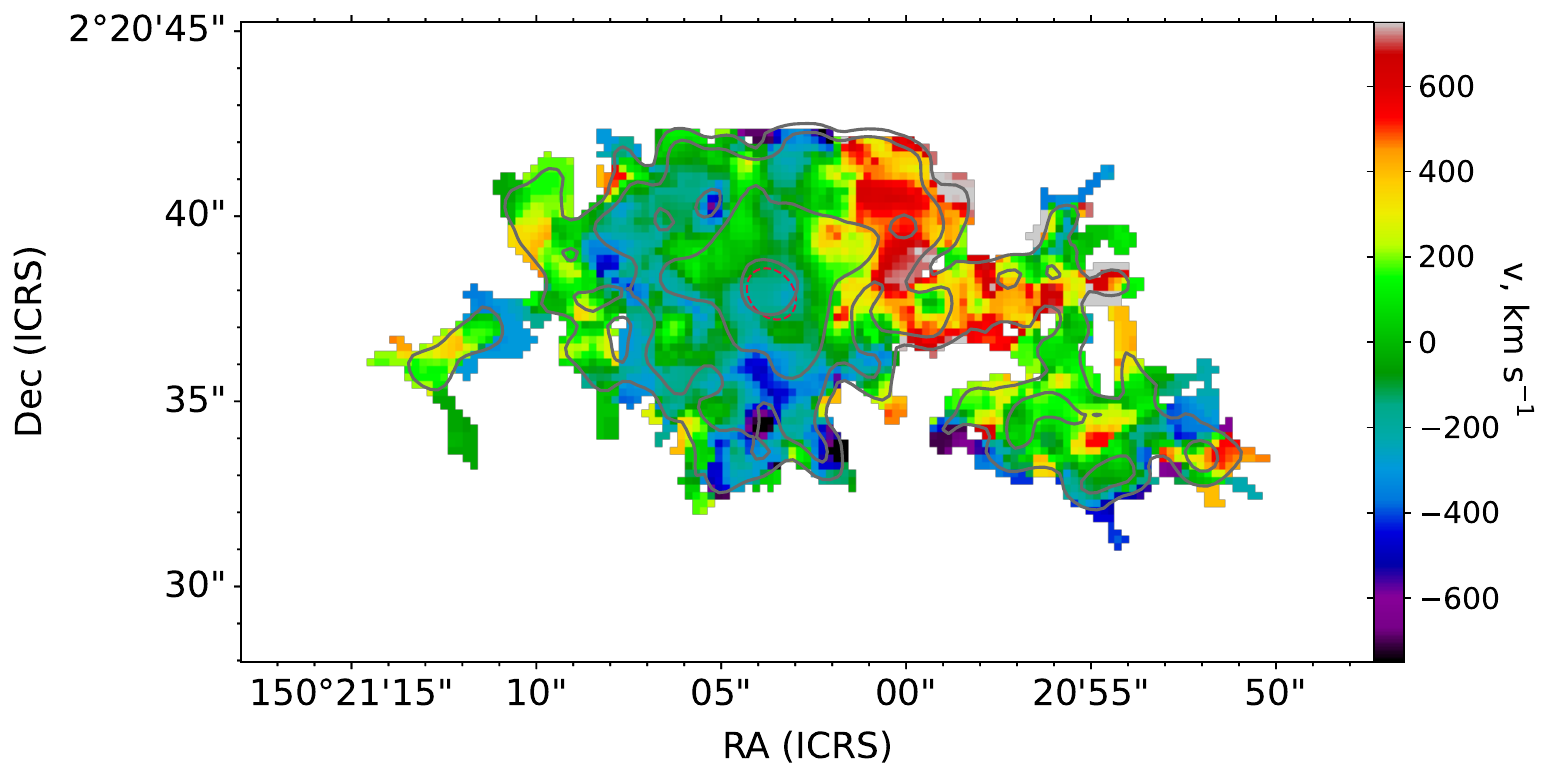}
        \label{fig:v1-ro1001sat}
        \vspace{-1.125cm}
    \end{subfigure}
    \begin{subfigure}{0.5\textwidth}
        \centering
        \includegraphics[width=\textwidth]{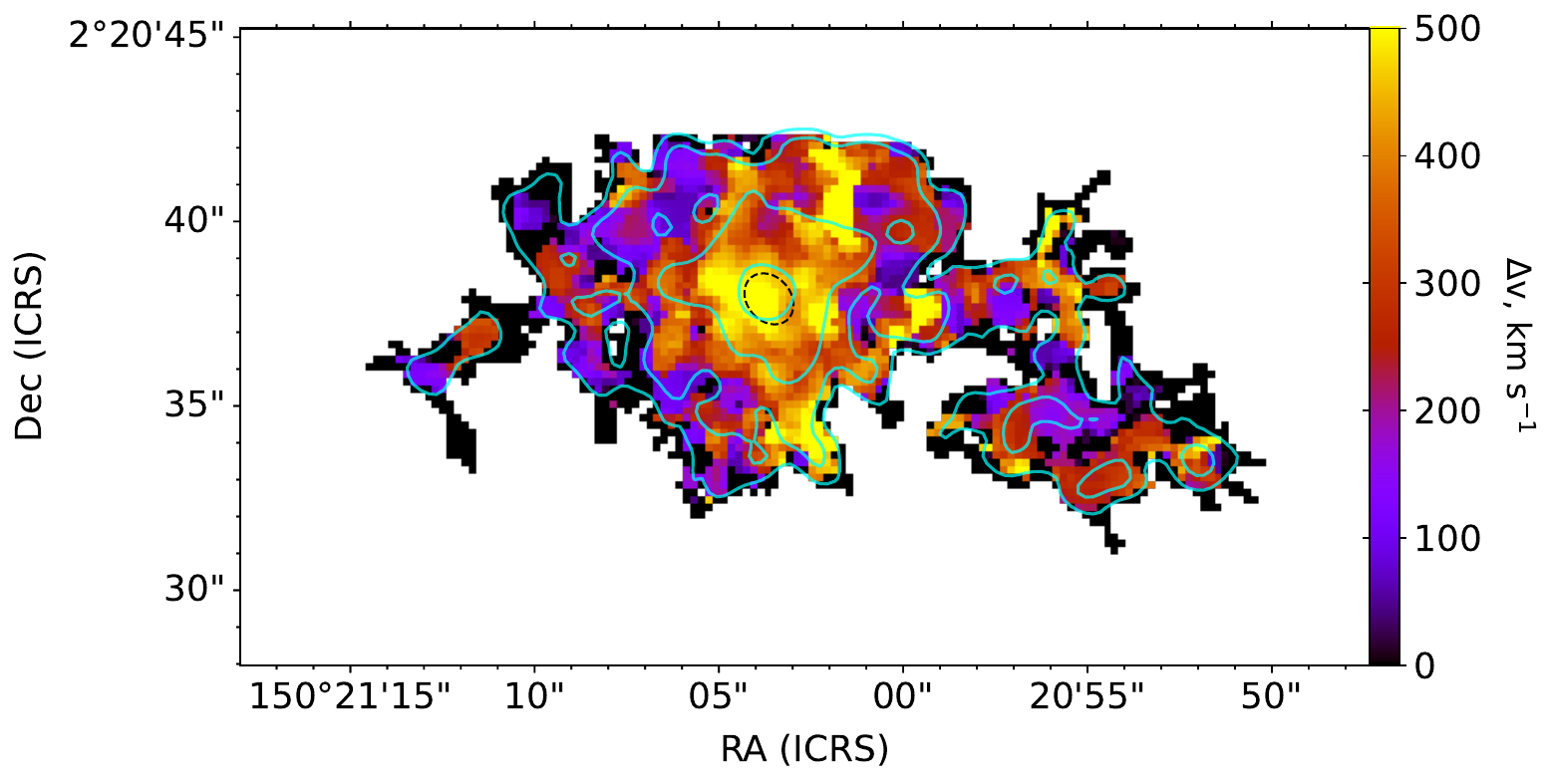}
        \label{fig:v2-ro1001sat}
        \vspace{-0.4cm}
    \end{subfigure}
    \caption{Ly$\alpha$ SB (\textit{top}), velocity (\textit{middle}) and velocity dispersion (\textit{bottom}) maps of RO-1001-Sat. The velocity map is referenced to the flux-weighted Ly$\alpha$ peak as the zero point. The contours in each map show the Ly$\alpha$ SB of the same levels as in Figure \ref{fig:rgb_ro1001}. The dashed ellipse at the centre of each map shows the location and approximate shape of the BGG as seen in Figure \ref{fig:rgb_ro1001}.}
    \label{fig:sb_v2_ro1001sat}
\end{figure}

\begin{figure}
    \centering
    \begin{subfigure}{0.272\textwidth}
        \centering
        \includegraphics[width=\textwidth]{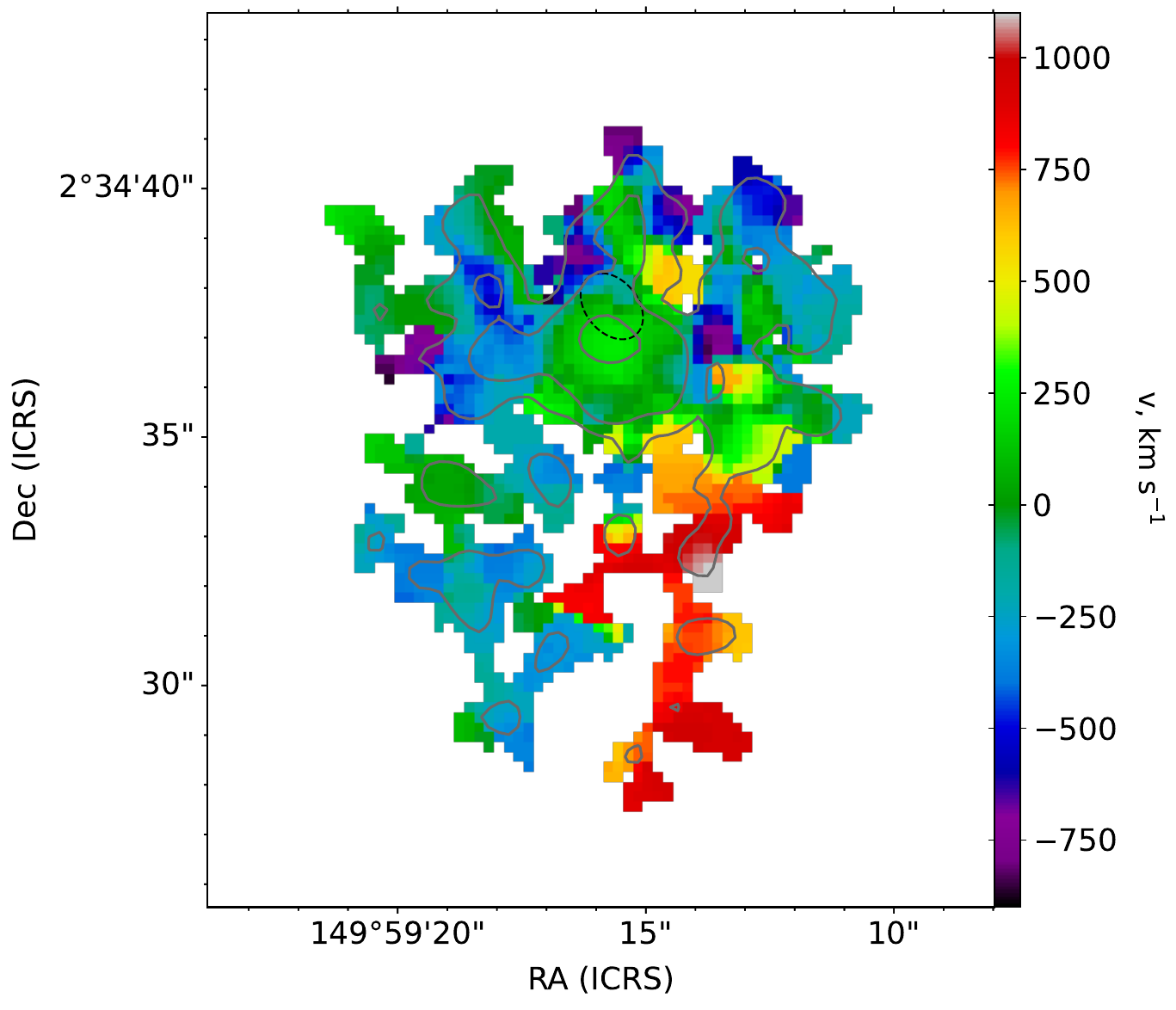}
        \label{fig:v2-ro0959sat}
    \end{subfigure}
    \hspace{-9.3cm}
    \begin{subfigure}{0.272\textwidth}
        \centering
        \includegraphics[width=\textwidth]{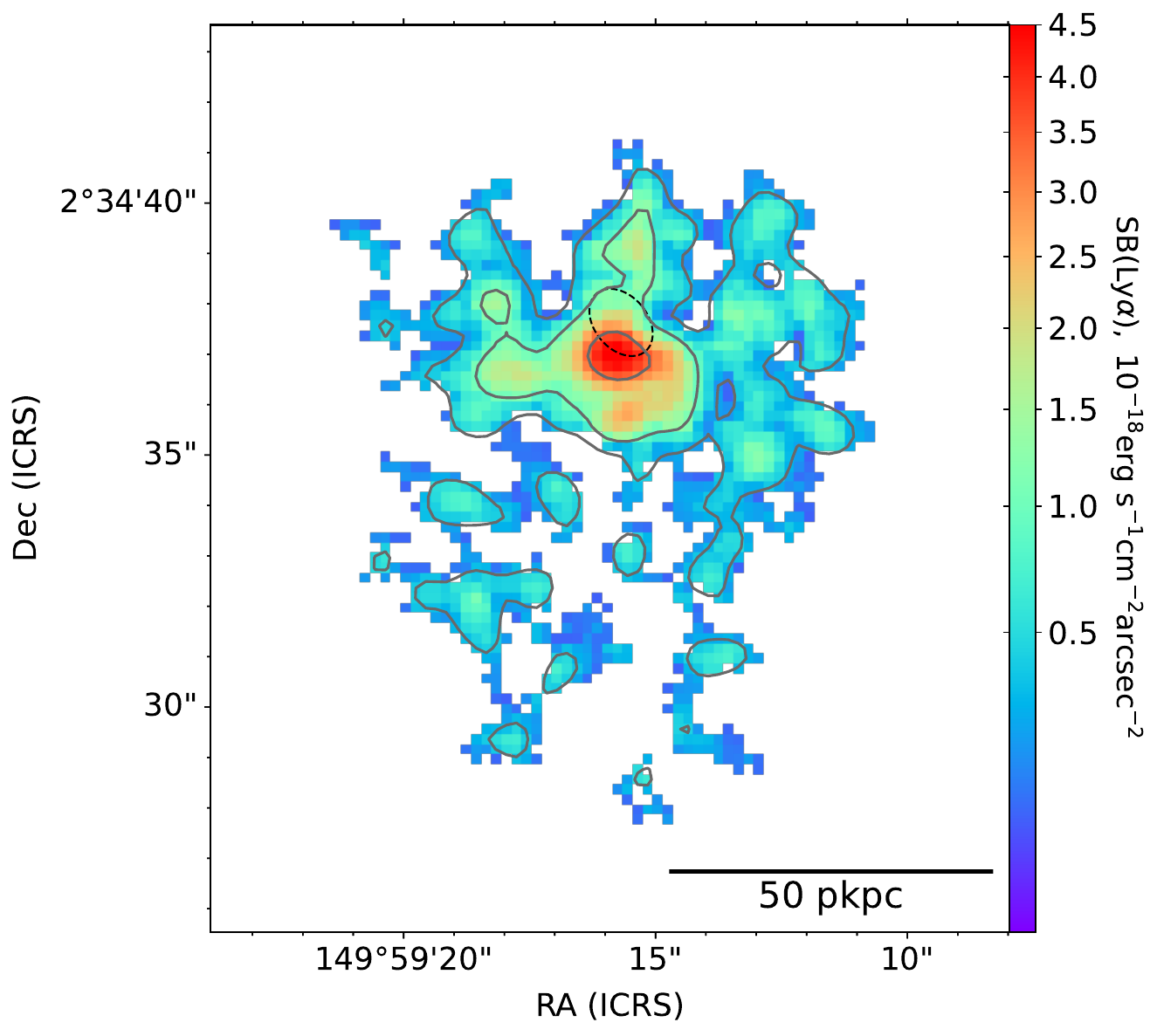}
        \label{fig:sb-ro0959sat}
    \end{subfigure}
    \hspace{5cm}
    \begin{subfigure}{0.272\textwidth}
        \vspace{-0.4cm}
        \centering
        \includegraphics[width=\textwidth]{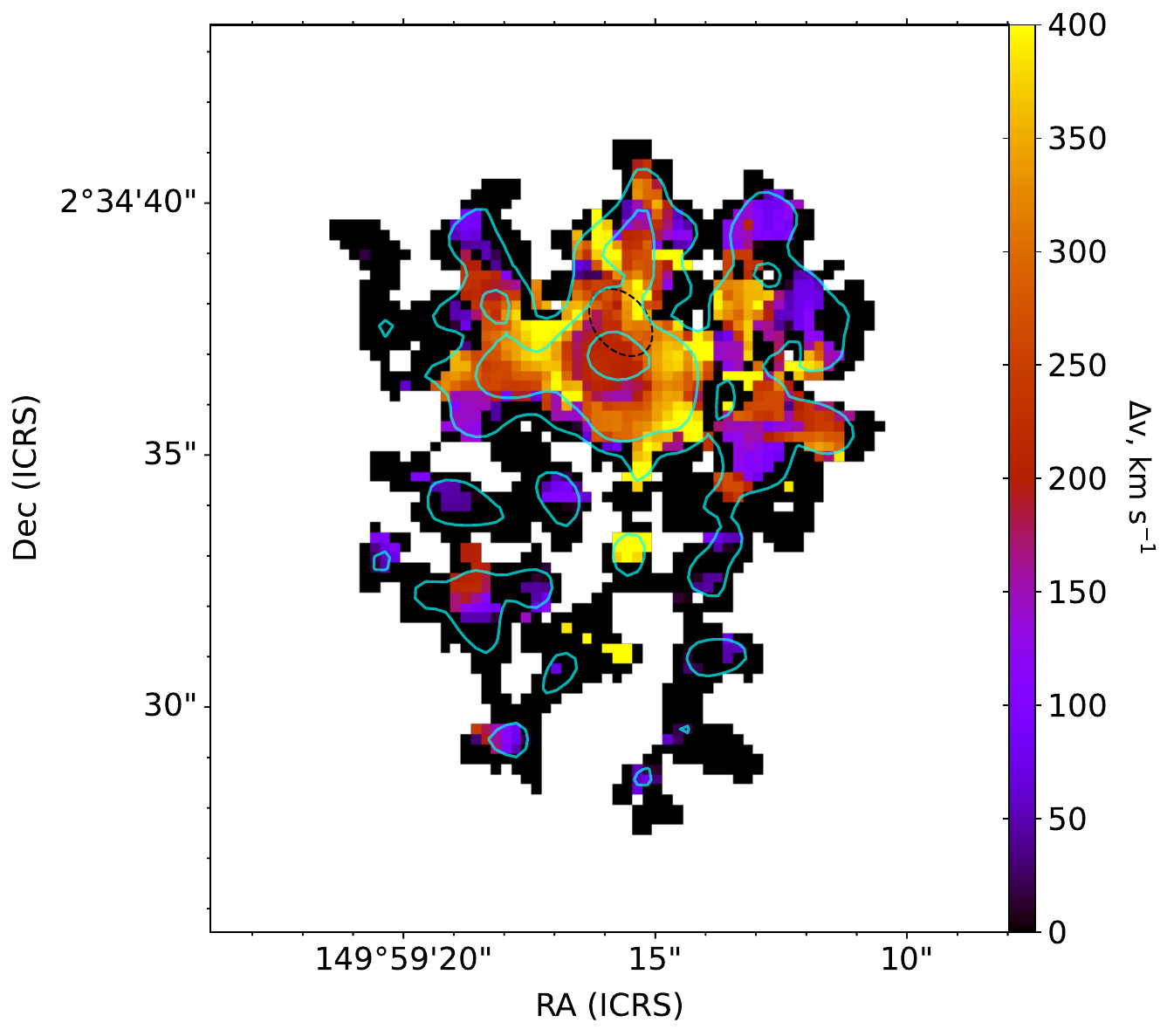}
        \label{fig:sb-ro0959sat}
    \end{subfigure}
    \vspace{-0.4cm}
    \caption{Ly$\alpha$ SB (\textit{top left}), velocity (\textit{top right}), and velocity dispersion (\textit{bottom}) maps of RO-0959-Sat. The velocity map is referenced to the flux-weighted Ly$\alpha$ peak as the zero point. The contours in each map show the Ly$\alpha$ SB of the same levels as in Fig. \ref{fig:rgb_ro0959}. The dashed ellipse at the centre of each map shows the location and approximate shape of the BGG as seen in Fig. \ref{fig:rgb_ro0959}.}
    \label{fig:sb_v2_ro0959sat}
\end{figure}

In the continuum-subtracted data cubes, we identify two smaller extended areas of emission at $\sim35-40\arcsec$ from the main nebulae of RO-1001 and RO-0959 (Figs. \ref{fig:rgb_ro1001} and \ref{fig:rgb_ro0959}), suggesting that they are at similar redshifts. This association is further strengthened by the redshifts of the central galaxies embedded in these two secondary nebulae. The central galaxy in RO-1001-Sat is photometrically estimated to be at $z=2.91^{+0.08}_{-0.44}$ in COSMOS2020 (LePhare), while the central galaxy in RO-0959-Sat has been spectroscopically confirmed at $z=3.091$ by CO(3-2) detection from ALMA observations (Daddi et al., in prep.). Compared to the two main nebulae, where the Ly$\alpha$ SB does not peak on any member galaxies, the Ly$\alpha$ peak aligns perfectly with the central galaxy in RO-1001-Sat and is only slightly shifted ($\simeq0.7\arcsec$) from the central galaxy in RO-0959-Sat. RO-1001-Sat appears to be at a projected distance of $\simeq$ 311 pkpc (1.22 cMpc), measured between the two Ly$\alpha$ peaks in the two nebulae, to the north-east of the main nebula and spans across $\simeq80\times$160 pkpc$^2$. It appears to be elongated in the east-west direction and has a tail pointing south-west, hinting that there could be a connection to the north-west filament of the main halo (see \citealt{Daddi21} for more on the three filaments in RO-1001). RO-0959-Sat resides to the west of the main nebula with a projected distance of $\simeq$ 276 pkpc (1.13 cMpc) measured between the Ly$\alpha$ peaks. It covers an area of $\simeq80\times$100 pkpc$^2$ extending to the south with filamentary morphology.

\subsubsection{Ly$\alpha$ spectra and kinematics} \label{sec:kinematics}
Fig. \ref{fig:lya_specs} shows the integrated Ly$\alpha$ spectra extracted from the four nebulae. We see that the flux-weighted peaks in the satellites only shift slightly (2-5$\AA$) from the main nebulae, equivalent to $\Delta v \simeq$ 304 km s$^{-1}$ and 123 km s$^{-1}$ in RO-1001 and RO-0959, respectively. This corresponds to redshifts of the two satellite structures being at $z=$ 2.920 for RO-1001-Sat and $z=$ 3.092 for RO-0959-Sat. By interpreting the redshifts solely by spatial displacement, the distance between RO-1001-Sat and its main nebula is calculated to be $\simeq$ 1 pMpc. However, assuming the relative velocity between such groups as their virial velocities, that are typically $\sim$500 km s$^{-1}$ as predicted by cold accretion models \citep{Goerdt10}, the uncertainty in the distance calculation can be large, giving an estimated distance between 0.3 - 2.8 pMpc. Similarly, the distance between RO-0959 and its satellite nebula is estimated to be $\simeq$ 0.4 pMpc and 0.3 - 2.0 pMpc if taking the relative velocity into account. The Ly$\alpha$ spectra from both satellite nebulae appear to be blueshifted with respect to the associated main LANs, hinting at gas accretion (e.g. \citealt{Dijkstra06, Verhamme06, Qiu20, Chen20, Daddi21}). Alternatively, the sharp jump on the blue side of the peak in the spectrum from RO-1001-Sat could also originate from absorption by foreground gas pertaining to the structure. The spectrum of RO-1001 also exhibits a blueshifted peak, that indicates gas infalls, as already discussed in \citet{Daddi21}. While the spectrum of RO-0959 shows a redshifted peak, that instead suggests outflows (e.g. \citealt{Verhamme06, Kulas12, Hashimoto13, Erb14}).

Figs. \ref{fig:sb_v2_ro1001sat} and \ref{fig:sb_v2_ro0959sat} show the Ly$\alpha$ SB, velocity, and velocity dispersion maps for the two satellite halos. In RO-1001-Sat, we see a steep jump from $-150\pm60~\mathrm{km~s}^{-1}$ at the centre to $530\pm100~\mathrm{km~s}^{-1}$ at the west side of the nebula, compared to the relatively flat velocity distribution towards the east side. This sharp jump could be due to gas accretion, AGN outflows, or spatial separations between gas clouds along the LOS. However, the lack of a redshifted component in the Ly$\alpha$ spectrum in Fig. \ref{fig:lya_specs} might disfavour the presence of AGN outflows. Meanwhile, the velocity dispersion peaking at $500\pm30~\mathrm{km~s}^{-1}$ at the lower end of the velocity gradient (i.e. the centre of the nebula) could provide further support to the scenario that the west tail might be a gas filament being accreted onto the centre of the potential well of the nebula, where the high gas turbulence reduces its velocity. On the contrary, the velocity dispersion distribution in RO-0959-Sat appears to peak at $340\pm50~\mathrm{km~s}^{-1}$ encircling the Ly$\alpha$ SB peak, while at the Ly$\alpha$ peak the dispersion instead goes down to $210\pm20~\mathrm{km~s}^{-1}$. The south-west filamentary structure also shows a high velocity of $800\pm140~\mathrm{km~s}^{-1}$, similar to the tail in RO-1001-Sat. This might be a similar scenario as in RO-1001-Sat, although the velocity gradient towards the centre is less smooth. However, it is worth noting that radiative transfer effects can alter the observed kinematics, which complicates the interpretation of the signatures.

\subsubsection{Halo mass estimates} \label{sec:Mh}
\begingroup
\renewcommand{\arraystretch}{1.4}
\begin{table*}
    \caption{\label{tab:member_galaxies}Candidate member galaxies identified from COSMOS2020 within the estimated virial radii of (a) RO-1001-Sat, and (b) RO-0959-Sat.}
    \begin{subtable}{\textwidth}
        \centering
        \begin{tabular}{>{\centering\arraybackslash}p{1.5cm} >{\centering\arraybackslash}p{2.2cm} >{\centering\arraybackslash}p{2.2cm} >{\centering\arraybackslash}p{1.8cm} >{\centering\arraybackslash}p{1.8cm} >{\centering\arraybackslash}p{1.5cm} >{\centering\arraybackslash}p{1.5cm}}
            \hline
            ID & R.A. & Dec. & $z_\mathrm{phot}$ & log$(M_*)~(M_\odot)$ & SFR ($M_\odot~\mathrm{yr^{-1}}$) & $d_{\mathrm{Ly}\alpha}$\tablefootmark{a} ($\arcsec$) \\ \hline
            976566 & 150.3509872 & 2.3438518 & $2.91^{+0.08}_{-0.44}$ & $11.3^{+0.2}_{-0.0}$ & $1^{+6}_{-0}$ & 0.2 \\
            973711 & 150.3504033 & 2.3423635 & $3.15^{+0.67}_{-0.23}$ & $10.1^{+0.1}_{-0.2}$ & $33^{+17}_{-25}$ & 5.9 \\
            974107 & 150.3498499 & 2.3427528 & $3.20^{+0.24}_{-0.22}$ & $9.6^{+0.1}_{-0.1}$ & $9^{+3}_{-3}$ & 5.9 \\
            977626 & 150.3508365 & 2.3455787 & $2.93^{+0.10}_{-0.13}$ & $9.5^{+0.1}_{-0.1}$ & $5^{+1}_{-1}$ & 6.1 \\
            973892 & 150.3505759 & 2.3420574 & $2.97^{+0.11}_{-0.09}$ & $9.6^{+0.1}_{-0.1}$ & $16^{+7}_{-12}$ & 6.8 \\
            975257 & 150.3491109 & 2.3437248 & $3.08^{+0.24}_{-0.25}$ & $9.7^{+0.1}_{-0.1}$ & $8^{+3}_{-2}$ & 6.9 \\
            976011 & 150.3533100 & 2.3437876 & $2.896^{+0.002}_{-0.002}$\tablefootmark{b} & $9.7^{+0.1}_{-0.1}$ & $56^{+11}_{-11}$ & 8.2 \\
            979058 & 150.3514126 & 2.3467909 & $3.22^{+0.12}_{-0.12}$ & $9.9^{+0.1}_{-0.1}$ & $20^{+4}_{-5}$ & 10.5 \\
            971930 & 150.3521540 & 2.3404616 & $3.18^{+0.14}_{-0.12}$ & $10.3^{+0.1}_{-0.1}$ & $22^{+17}_{-7}$ & 13.0 \\
            981015 & 150.3491015 & 2.3489236 & $2.84^{+0.46}_{-0.40}$ & $10.1^{+0.1}_{-0.2}$ & $24^{+18}_{-13}$ & 19.4 \\
            970048 & 150.3527952 & 2.3386652 & $2.69^{+0.09}_{-0.06}$ & $9.8^{+0.1}_{-0.1}$ & $14^{+13}_{-3}$ & 19.9 \\
            981411 & 150.3485605 & 2.3490356 & $2.81^{+0.18}_{-0.16}$ & $9.7^{+0.1}_{-0.2}$ & $13^{+11}_{-3}$ & 20.5 \\ \hline
        \end{tabular}
        \caption{}
        \label{tab:galaxies_ro1001}
    \end{subtable}
    \begin{subtable}{\textwidth}
        \centering
        \begin{tabular}{>{\centering\arraybackslash}p{1.5cm} >{\centering\arraybackslash}p{2.2cm} >{\centering\arraybackslash}p{2.2cm} >{\centering\arraybackslash}p{1.8cm} >{\centering\arraybackslash}p{1.8cm} >{\centering\arraybackslash}p{1.5cm} >{\centering\arraybackslash}p{1.5cm}}
            \hline
            ID & R.A. & Dec. & $z_\mathrm{phot}$ & log$(M_*)~(M_\odot)$ & SFR ($M_\odot~\mathrm{yr^{-1}}$) & $d_{\mathrm{Ly}\alpha}$ ($\arcsec$) \\ \hline
            1220803 & 149.9876525 & 2.5771193 & 3.091\tablefootmark{c} & $11.2^{+0.1}_{-0.1}$ & $175^{+42}_{-142}$ & 0.7 \\
            1219049 & 149.9844740 & 2.5765188 & $3.21^{+0.10}_{-0.09}$ & $9.9^{+0.1}_{-0.1}$ & $23^{+5}_{-9}$ & 11.6 \\ \hline
        \end{tabular}
        \caption{}
        \label{tab:galaxies_RO0959}
    \end{subtable}
    \tablefoot{\tablefoottext{a}{Distance to the peak of the LAN.} \tablefoottext{b}{The redshift of this galaxy has been spectroscopically confirmed from its MUSE spectrum.} \tablefoottext{c}{The redshift of this galaxy has been spectroscopically confirmed by the detection of CO(3-2) from ALMA observations.}}
\end{table*}
\endgroup

Because diffuse Ly$\alpha$ emission has been proposed as a tracer of massive halos at high $z$ \citep{Daddi22a}, we consider these two satellite structures to be two candidate galaxy groups. Therefore, to identify potential group members and to estimate the hosting halo mass, we applied the same methods as in \cite{Sillassen24} by iteratively selecting candidate members from COSMOS2020 within the estimated projected virial radii, $R_\mathrm{vir}$, of the satellite halos and with $|z_\mathrm{phot}-z_\mathrm{spec,Ly\alpha}|<0.1(1+z_\mathrm{spec,Ly\alpha})$, where $z_\mathrm{spec,Ly\alpha}$ is the redshift of the two satellite LANs in Table \ref{tab:obs_details}. Following the halo mass estimation methods 2a, 3, and 4 presented in \citet{Sillassen24}, we first calculated the total stellar mass above the QG completeness limit in COSMOS2020 from \cite{Weaver22} ($\log M_\ast/{\rm M_\odot}>9.5$, at $z\sim3$, 70\%) to minimise the number of interloper galaxies, extrapolated it down to $M_\ast=10^7\,{\rm M_\odot}$ assuming the field stellar mass function from \citet{Muzzin13}, and corrected it for the background contribution. The background-corrected total stellar mass was scaled to the stellar to halo mass relation (SHMR) from \citet{Shuntov22}. This gives estimations of log$(M_{\rm h}/M_\odot)=13.0^{+0.3}_{-0.2}$ and $12.9^{+0.3}_{-0.2}$ for RO-1001-Sat and RO-0959-Sat, respectively. Alternatively, we calculated the average number density of the group candidates selected with the criteria as stated above (within $R_\mathrm{vir}$ and in the redshift range) after applying a stellar mass cut so that the interloper fraction at $R_\mathrm{vir}$ was $<10\%$. We estimated the overdensity of the satellites compared to the field using the method in \cite{Sillassen22}, where the field density is defined as the average number density of the mass-complete redshift-selected galaxies across the COSMOS field. We then iteratively calculated the clustering bias (as $\simeq$12.68 for RO-1001-Sat and $\simeq$8.34 for RO-0959-Sat) using the \citet{Tinker10} formalism with the SHMR method above as an initial guess, and used Eq. (3) of \citet{Sillassen24} to estimate the halo mass. This method yields log$(M_{\rm h}/M_\odot)= 13.1^{+0.2}_{-0.3}$ for RO-1001-Sat and log$(M_{\rm h}/M_\odot)= 12.8^{+0.1}_{-0.2}$ for RO-0959-Sat. Finally, since there are more identified candidate members in RO-1001-Sat, we made a third estimate of its halo mass by fitting the radial stellar mass density profile to a projected Navarro-Frenk-White \citep{Navarro1997} model, which gives log$(M_{\rm h}/M_\odot)= 13.5^{+0.2}_{-0.3}$. Since the estimates from all the methods above are consistent, we adopted an average of the three or two methods as the reference halo masses, log$(M_{\rm h}/M_\odot)=13.2\pm0.3$ for RO-1001-Sat and log$(M_{\rm h}/M_\odot)=12.8\pm0.3$ for RO-0959-Sat, respectively. 

The candidate member galaxies identified are listed in Table \ref{tab:member_galaxies}. 12 candidate member galaxies with a total stellar mass of $M_*\simeq10^{11.4}M_\odot$ are identified within the estimated $R_\mathrm{vir}$ of RO-1001-Sat, one of which is spectroscopically confirmed as noted in Table \ref{tab:galaxies_ro1001} by fitting its MUSE spectrum with \texttt{czspecfit} (Gobat et al., in prep.). Using the same methods as in \cite{Sillassen24}, the interloper fraction is expected to be 32.7\% by number and 29.5\% by mass. This suggests that 4$\pm$2 out of 12 galaxies with a total $M_*\simeq10^{10.9}M_\odot$ could be spurious, and we expect most of them likely to be the least massive ones. If we limit the aperture size to 10$\arcsec$, the interloper fractions drop to 8.6\% ($\simeq$1) by number and 4.9\% ($\simeq10^{10.1}M_\odot$) by mass. However, only considering candidates inside 10$\arcsec$ does not affect the estimates significantly, as most of the $M_\mathrm{h}$ is driven by the central massive galaxy. As for RO-0959-Sat, the central galaxy has been spectroscopically confirmed by CO(3-2) emission from ALMA observations (Daddi et al., in prep). Only one more candidate member galaxy is identified to be within the $R_\mathrm{vir}$. There could be fainter member galaxies below the detection limit that are not identified by our method, which should also be the case for RO-1001-Sat. Although only two (candidate) member galaxies are identified to be within its $R_\mathrm{vir}$, RO-0959-Sat is still estimated to be fairly massive due to the high total $M_*$ that is mainly concentrated on the central galaxy. RO-1001-Sat is estimated to be 0.4$\pm$0.3 dex less massive than the main halo, while RO-0959-Sat appears to be approximately as massive as its main halo, with a similar uncertainty.

We derive the virial radius from the estimated halo mass using the $M_\mathrm{h}-R_\mathrm{vir}$ relation in \citet{Goerdt10} to be $R_{\rm vir}=197\pm45$ kpc $\simeq25\arcsec$ for RO-1001-Sat and $139\pm32$ kpc $\simeq18\arcsec$ for RO-0959-Sat. While for the main halos, $R_{\rm vir}$ are estimated to be $269\pm41$ and $139\pm32$ kpc for RO-1001 and RO-0959 using their $M_\mathrm{h}$ from Table \ref{tab:obs_details}. We see that the estimated halo sizes of RO-1001 and RO-1001-Sat are overlapping in the projected plane, as shown in Fig. \ref{fig:rgb_ro1001}. Along the LOS, however, they could be much further apart due to the large uncertainty in the estimated distance discussed above. On the other hand, $R_\mathrm{vir}$ of RO-0959 and RO-0950-Sat appear to be adjacent in the plane of sight with identical sizes due to their comparable $M_\mathrm{h}$. However, $R_\mathrm{vir}$ of RO-0959 seems to be unexpectedly smaller than the other three nebulae compared to the LAN expanse, which suggests that its $M_\mathrm{h}$ might have been underestimated. We measured the Ly$\alpha$ luminosity from the narrow-band images of the two satellite nebulae as log$(L_{\mathrm{Ly}\alpha}/\mathrm{erg~s}^{-1})=43.4$ for RO-1001-Sat, 0.6 dex lower than the main halo, and log$(L_{\mathrm{Ly}\alpha}/\mathrm{erg~s}^{-1})=42.9$ for RO-0959-Sat, 1.1 dex lower than its main halo. The uncertainties of both $L_{\mathrm{Ly}\alpha}$ measurements are $<$0.01 dex.

\subsection{Near neighbour probability} \label{sec:near_neighbour_prob}
\begin{figure}
    \centering
    \includegraphics[width=\linewidth]{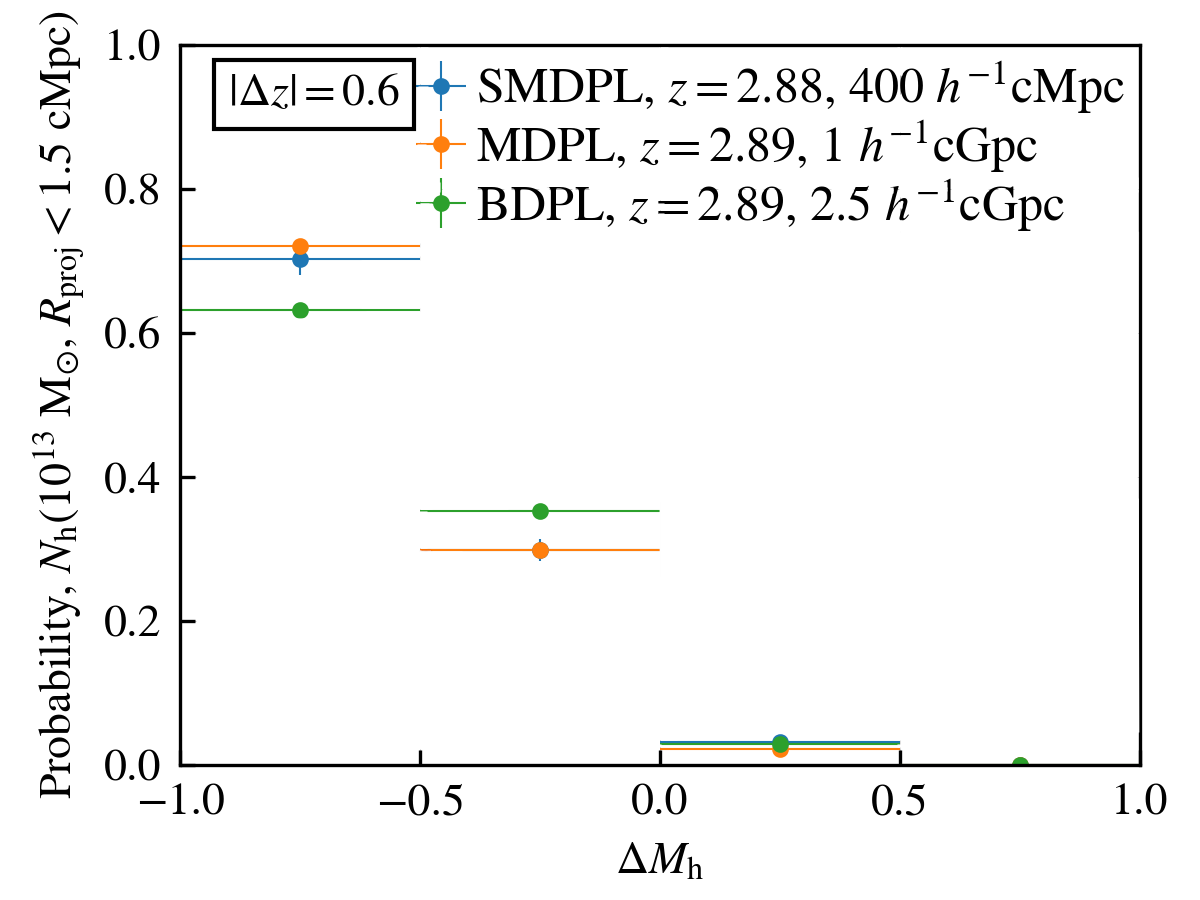}
    \caption{Probabilities of finding a neighbouring halo with log$(M_\mathrm{h}/M_\odot)$ = (12-12.5), (12.5-13), (13-13.5) and (13.5-14) within 1.5 cMpc projected distance from a central massive halo of log$(M_\mathrm{h}/M_\odot) = 12.7-13.3$ in the MultiDark simulations. Integrated probabilities within $\Delta z$ = 0.6 from the SMDPL, MDPL and BMDPL simulations are shown in the blue, orange and green points.}
    \label{fig:prob_vs_delMh}
\end{figure}

Since there have not been similar close-by massive halo pairs at $z\gtrsim3$ explicitly identified previously, we seek to cross-check with simulations to inspect whether this would be reasonably expected given the current cosmological framework, and also whether the halo masses we estimated are not unplausible. To examine how likely it is to find such a massive satellite halo in the vicinity of a massive main halo at $z\sim3$ in simulations, we looked into the MultiDark simulation suite (\citealt{Riebe13, Klypin16}) and made use of the Small MultiDark Planck simulation (SMDPL), MultiDark Planck simulation (MDPL), and Big MultiDark Planck simulation (BMDPL) with co-moving volumes of (400 $h^{-1}$ Mpc)$^{3}$, (1 $h^{-1}$ Gpc)$^{3}$ and (2.5 $h^{-1}$ Gpc)$^{3}$, respectively. See Appendix \ref{app:multidark} for more details of the simulations and the valid halo mass ranges for the probability estimates.

Based on a sample of halos with masses in the range of log$(M_\mathrm{h}/M_\odot)=12.7-13.3$, we searched for the nearest neighbours using KDTree\footnote{\url{https://docs.scipy.org/doc/scipy/reference/generated/scipy.spatial.KDTree.html}} with log$(M_\mathrm{h}/M_\odot)=13\pm\Delta M_\mathrm{h}$ where $\Delta M_\mathrm{h}$ ranges from -1 to +1 dex. The probabilities of finding a massive neighbour with the halo mass being log$(M_\mathrm{h}/M_\odot)=(12-12.5),~(12.5-13.0),~(13-13.5)~\mathrm{and}~(13.5-14.0)$ within a 1.5 cMpc projected distance from the main halo are shown in Fig. \ref{fig:prob_vs_delMh}. The blue, orange and green points correspond to the SMDPL, MDPL, and BMDPL simulations. The probabilities shown are integrated within a redshift range of $\Delta z~\pm$ 0.6. Due to the large uncertainties in the distances between the two LAN pairs as estimated in Sect. \ref{sec:satellites}, we chose this conservative redshift range, that is large enough to contain all physical clustering signals, while any signal beyond this scale is negligible. Nevertheless, considering the extremely low probability of finding two overlapping yet uncorrelated halos within the mass range concerned (\citealt{Sillassen24}), almost all the massive neighbours found are expected to be close to the main halos. The figure shows that the probability increases with decreasing neighbouring halo mass. This is expected since massive halos are prone to having smaller structures in the surroundings according to the hierarchical formation theory. For a host halo of $\simeq10^{13}\rm{M}_{\odot}$, we estimate the probability of finding a neighbouring halo in the mass range of log$(M_\mathrm{h}/M_\odot)=(12.5-13.0)$ within a projected aperture of 1.5 cMpc to be $\simeq$ 30-35.5\%. We observe that the simulations do not converge well, likely due to the limited halo clustering in the volume of each simulation. 

In our sample, we find two satellite halos in the mass range of log$(M_\mathrm{h}/M_\odot)=(12.5-13.5)$ around four massive host halos with log$(M_\mathrm{h}/M_\odot)=(12.8-13.6)$ (here we exclude the galaxy group where the Ly$\alpha$ emission line is not covered by MUSE from our sample). In comparison, using the BDPL simulation that has the largest volume, the average probability of finding at least one satellite halo with log$(M_\mathrm{h}/M_\odot)=(12.5-13.0)$ within a projected distance of 1.5 cMpc from a main halo with log$(M_\mathrm{h}/M_\odot)=13.0$ is estimated to be $\simeq 76\%$ assuming a Poisson probability distribution. We also find that the probability of finding a neighbouring halo decreases with increasing host halo mass, likely due to the low occurrence of massive halos at $z\sim3$ in the simulations. Hence, in practice, this estimated probability (76\%) should be lower for more massive halos such as RO-1001. Nevertheless, the result from the simulations is consistent with our small number statistics.

\subsection{Brightest group galaxies (BGGs)} \label{sec:BGGs}
\begin{figure}
    \centering
    \begin{subfigure}{\linewidth}
        \centering
        \includegraphics[width=\textwidth]{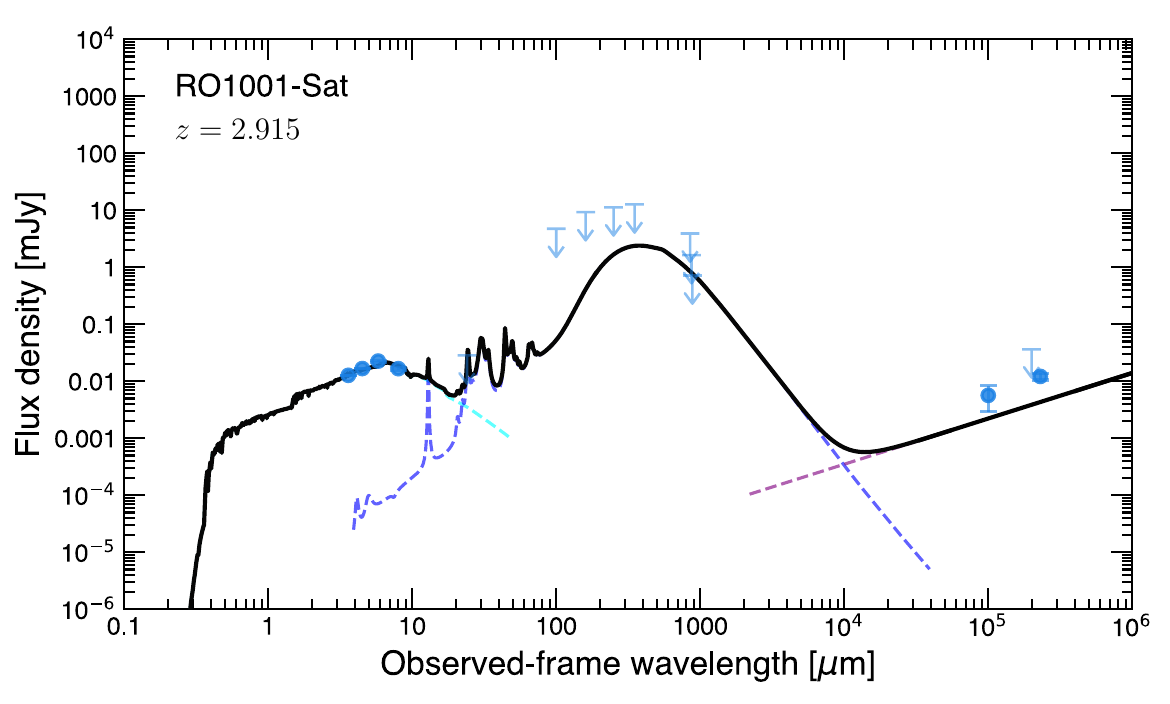}
        \label{fig:sed-ro1001-fir}
    \end{subfigure}
    \begin{subfigure}{\linewidth}
        \centering
        \includegraphics[width=\textwidth]{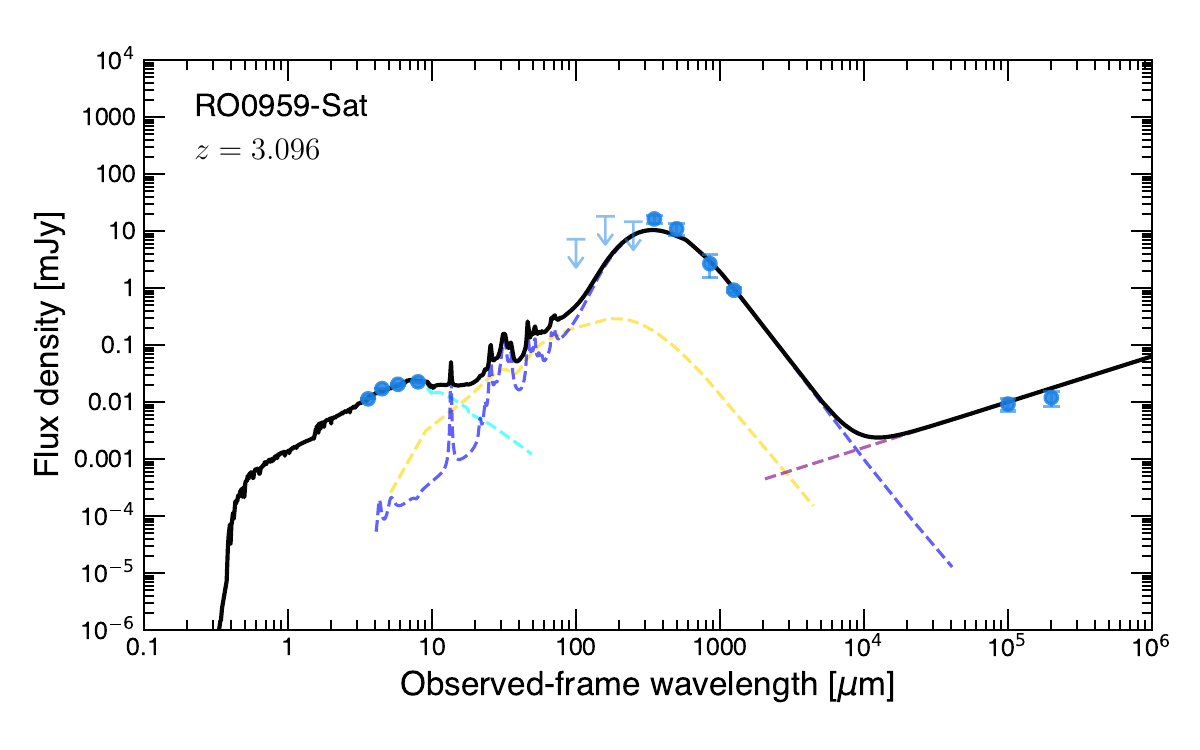}
        \label{fig:sed-ro0959-fir}
    \end{subfigure}
    \caption{Best-fit SEDs (black curves) in FIR-to-radio for (\textit{top}) RO-1001-Sat and (\textit{bottom}) RO-0959-Sat. The SED components shown in dashed lines are stellar (cyan), mid-IR AGN (yellow), cold dust (blue), and radio (purple). Photometry data from COSMOS2020, JWST, ALMA archive and the latest super-deblended catalogue are shown in circles or upper limits.}
    \label{fig:seds-fir}
\end{figure}

\begin{table}
    \caption{\label{tab:sed_params}Physical parameters of the two satellite nebulae derived from FIR-to-radio SEDs and X-ray data.}
    \begin{subtable}{0.5\textwidth}
        \centering
        \begin{tabular}{c c}
            \hline
            $L_\mathrm{AGN,~IR}~(\mathrm{erg~s}^{-1})$ & $\lesssim7.6\times10^{44}$\\
            $L_\mathrm{AGN,X,bol}~(\mathrm{erg~s}^{-1})$ & $\lesssim7.9\times10^{44}$\\
            log($L_\mathrm{IR, total}, L_\odot$) & $\lesssim$ 12.2\\
            log($M_\mathrm{dust, total}, M_\odot$) & $\lesssim$ 8.61\\
            $\langle U\rangle_\mathrm{total}$\tablefootmark{a} & -\\
            \hline
        \end{tabular}
        \caption{RO-1001-Sat}
        \label{tab:sed_params_ro1001}
    \end{subtable}
    \begin{subtable}{0.5\textwidth}
        \centering
        \begin{tabular}{c c}
            \hline
            $L_\mathrm{AGN,~IR}~(\mathrm{erg~s}^{-1})$ & $\lesssim4.8\times10^{45}$\\
            $L_\mathrm{AGN,X,bol}~(\mathrm{erg~s}^{-1})$ & $\lesssim1.2\times10^{45}$\\
            log($L_\mathrm{IR, total}, L_\odot$) & 12.48 $\pm$ 0.08\\
            log($M_\mathrm{dust, total}, M_\odot$) & 8.91 $\pm$ 0.12\\
            $\langle U\rangle_\mathrm{total}$ & 29.0 $\pm$ 11.4\\
            \hline
        \end{tabular}
        \caption{RO-0959-Sat}
        \label{tab:sed_params_ro0959}
    \end{subtable}
    \tablefoot{The FIR-to-radio SEDs are shown in Fig. \ref{fig:seds-fir}. $L_\mathrm{AGN,X,bol}$ was derived from X-ray data from the COSMOS Legacy survey (\citealt{Civano16, Marchesi16}) assuming the bolometric corrections from \citet{Lusso12} for type-2 AGN. \tablefoottext{a}{Mean interstellar radiation field intensity. $\langle U\rangle_\mathrm{total}$ is not reported for RO-1001-Sat as the SED fitting with only upper limits in IR is not constraining enough for a robust estimate.}}
\end{table}

\begin{figure}
    \centering
    \begin{subfigure}{\linewidth}
        \centering
        \includegraphics[width=\textwidth]{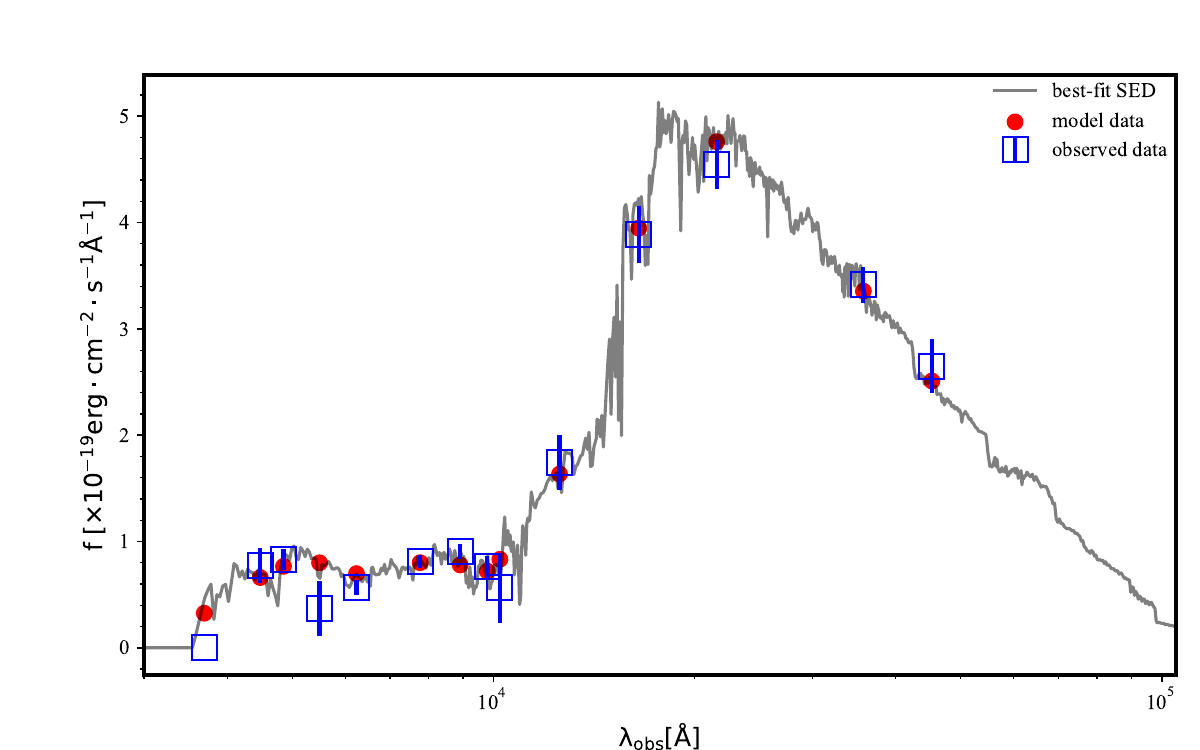}
        \label{fig:sed-ro1001-cosmos}
    \end{subfigure}
    \begin{subfigure}{\linewidth}
        \centering
        \includegraphics[width=\textwidth]{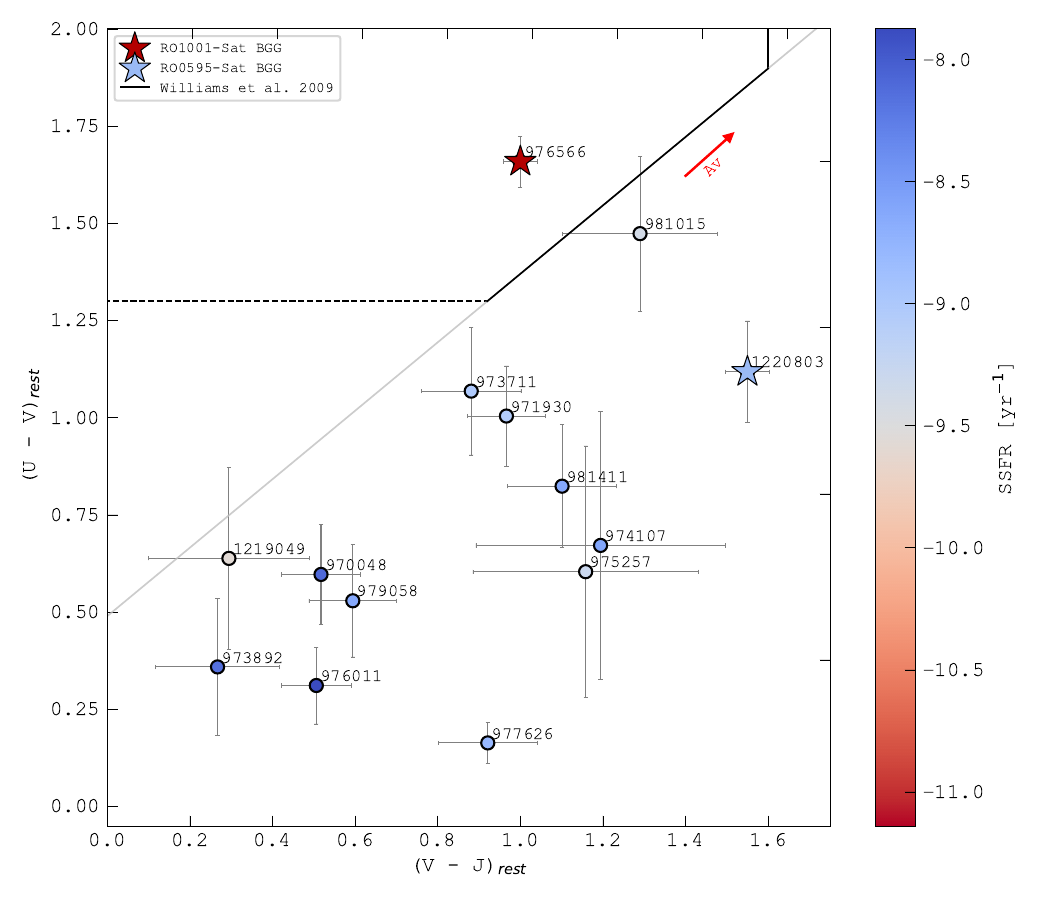}
        \label{fig:uvj-ro1001}
    \end{subfigure}
    \caption{(a) Best-fit optical SED (grey curve) of the BGG in RO-1001-Sat using data (red points) from COSMOS2020. (b) UVJ colours of all candidate group members of RO-1001-Sat and RO-0959-Sat using their $z_\mathrm{phot}$ from COSMOS2020. The BGGs in RO-1001-Sat and RO-0959-Sat are denoted by the red and blue stars, respectively.}
    \label{fig:uvj_sed_cosmos}
\end{figure}

To study star formation and AGN activities in the two central galaxies in RO-1001-Sat and RO-0959-Sat, we constrained their far-infrared (FIR) to radio spectral energy distribution (SED) properties, utilising their FIR-to-radio photometry in the latest Super-deblended catalogue (\citealt{Jin18}; in prep.), spanning MIPS/24$\mu$m, Herschel, SCUBA-2, and radio bands, with additional IRAC and ancillary ALMA photometry where available. We ran a panchromatic SED fitting with the MiChi2 code \citep{Liu2021MiChi2}, adopting $z_\mathrm{spec}$ from the Ly$\alpha$ emission. In Michi2, the modelling of the radio component is tied to the FIR through the assumed IR-radio correlation (e.g. \citealt{Yun01, Ivison10}). Given that we only have upper limits in the FIR for the brightest group galaxy (BGG) in RO-1001-Sat, we performed the fitting without the radio data points for this galaxy to prevent the fit from being artificially driven by the radio component. The best-fit SEDs for the two galaxies are presented in Fig. \ref{fig:seds-fir}, and the derived physical parameters are listed in Table \ref{tab:sed_params}.

\subsubsection{A quiescent BGG in RO-1001-Sat} \label{sec:BGG_ro1001}
\begin{figure*}
    \centering
    \begin{subfigure}{0.49\linewidth}
        \centering
        \includegraphics[width=0.8\textwidth]{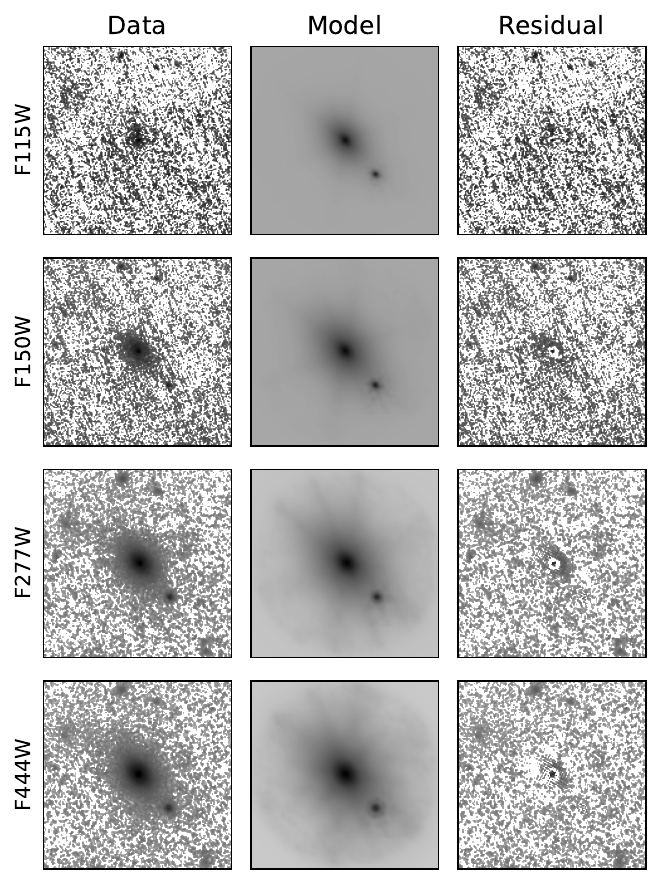}
        \caption{RO-1001-Sat}
        \label{fig:sersic_fitting/RO1001}
    \end{subfigure}
    \begin{subfigure}{0.49\linewidth}
        \centering
        \includegraphics[width=0.8\textwidth]{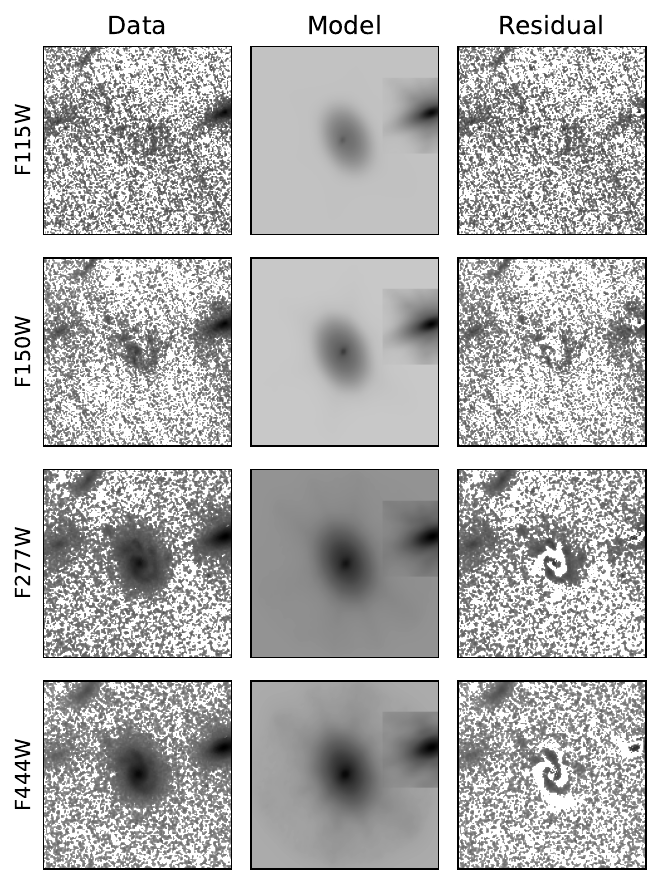}
        \caption{RO-0959-Sat}
        \label{fig:sersic_fitting/RO0959}
    \end{subfigure}
    \caption{Sérsic profile fitting for the BGGs of (a) RO-1001-Sat and (b) RO-0959-Sat. Each row shows the fit for a NIRCam filterband, from top to bottom: F115W, F150W, F277W, and F444W. The first column shows the data as 6$\arcsec\times6\arcsec$ cutouts centred on the respective BGG. The second column shows the best-fit models and the third column shows the residual images (Data-Model). The data, models, and residuals are displayed with the same logarithmic intensity scaling for each filterband.}
    \label{fig:sersic_fitting}
\end{figure*}

Due to the non-detection in FIR, the best-fit FIR-to-radio SED of the BGG in RO-1001-Sat yields a negligible AGN component and an infrared luminosity upper limit of log$(L_\mathrm{IR,total}/L_\odot) \lesssim 12.2$, which gives SFR$_\mathrm{IR}~\lesssim 158~M_\odot$ yr$^{-1}$ following the relation in \cite{Kennicutt98} (with a Chabrier IMF, i.e. SFR $\simeq10^{-10}L_\mathrm{IR}$). In contrast, this galaxy shows a clear radio excess, three times above the best-fit SED from the IR-radio correlation. Taking the spectral index from 1.4 - 3 GHz as $\alpha$ = -1.0 and the 1.4 GHz flux as 12.19 $\pm$ 1.66 mJy, we calculate the 1.4 GHz rest-frame luminosity as log$(L_\mathrm{1.4GHz}/ \mathrm{W~Hz}^{-1})=23.94\pm0.05$, which yields log$(L_{\rm IR, total}/ L_\odot) = 12.47$, and hence SFR$_\mathrm{radio}~\simeq 300~M_\odot$ yr$^{-1}$ \citep{Kennicutt98} instead, resulting in a $\sim2.7\sigma$ discrepancy from the SED fitting result.

We then fitted the optical to near-infrared (NIR) SED shown in the upper panel of Fig. \ref{fig:uvj_sed_cosmos} using FAST++\footnote{\url{https://github.com/cschreib/fastpp?tab=readme-ov-file}} and the photometry from CFHT-u, SC-B and V, HSC-g, r, i, z, and y, UVISTA-Y, J, H, and Ks, and IRAC-channel 1 and 2, corrected for the aperture correction and galactic extinction. We used the stellar templates by \cite{Maraston13} (M13) and the exponentially declining star formation histories (SFHs), as the M13 stellar templates accurately implement the thermally pulsating asymptotic giant branch stars, that contribute to a large fraction of the infrared emission of old stellar population \citep{Lu25}, while the exponentially declining SFHs are more fit to physically represent galaxies that have stopped forming stars or are in the process of doing so. The optical SED yields an SFR$_\mathrm{optical}~=2.51^{+0.50}_{-0.06}~M_\odot$ yr$^{-1}$ averaged over 50-100 Myr, showing an even larger discrepancy from the SFR derived from the radio luminosity. For easier comparisons to the literature, we used the stellar templates by \cite{Bruzual03} (BC03) to derive the age of the BGG as $1.26^{+0.12}_{-0.24}$ Gyr and the stellar mass as log$(M_*/M_\odot)=11.23^{+0.07}_{-0.02}$, and thus log(sSFR/yr$^{-1})=-10.83^{+0.08}_{-0.08}$. Using M13 does not significantly change the results.

The UVJ diagram in the lower panel of Fig. \ref{fig:uvj_sed_cosmos} shows that this BGG stands out from all the other candidate group members in RO-1001-Sat and the BGG in RO-0959-Sat to be the only galaxy residing above the black lines, that is the regime of QGs, although contaminations by dusty star-forming galaxies cannot be excluded either. In terms of the degeneracy in optical SEDs between the galaxy being quiescent and dust obscured, the archived ALMA data (project: 2013.1.00884.S, P.I.: David Alexander) that had this BGG covered at the edge of the FOV, where there is no detection (0.06 $\pm$ 0.24 mJy) at 340GHz, disfavours the dust obscured scenario. Therefore, based on the results from the well-fitted optical SED ($\chi^2\sim1$) as well as the lack of detection in ALMA, this BGG is classified as a QG. In addition, this galaxy has a stellar mass of log$(M_*/M_\odot)=11.24^{+0.10}_{-0.04}$ and log(sSFR/yr$^{-1})=-10.37^{+0.09}_{-0.30}$ in the COSMOS-WEB catalogue (BC03; \citealt{Shuntov25}), consistent with our SED fitting results and with it being quiescent. 

To analyse the morphology of the BGG, we fitted its SB profile with a 2D Sérsic profile \citep{Sersic68} using \textsc{AstroPhot} \citep{Stone23}. We fitted a single Sérsic profile in cutouts of $6\arcsec\times6\arcsec$ centred on the BGG in JWST NIRCam F115W, F150W, F277W and F444W bands simultaneously while keeping the Sérsic index, the axis ratio, the position angle and the effective radius constant over all bands, as using different values for different bands does not have a significant impact on the fitting results. A second offset Sérsic profile was fitted simultaneously to account for a nearby bright source, and a flat sky model was added as well. Fig. \ref{fig:sersic_fitting} (left) shows the data, the best-fit models, and the residuals. We find a Sérsic index of $n = 2.29\pm0.02$ and an effective radius of $r_{\text{eff}} = 1.98\pm0.01$ kpc. Fitting with an additional Sérsic profile to account for a possible bulge+disc system did not improve the fit significantly or change the results within the uncertainties. The Sérsic index is lower than the median found for massive QGs around $z\sim3$ (see \citealt{Kawinwanichakij25} and references therein), but does not clearly point to a disc-dominated structure either. The size is in agreement with what has been found in the literature for massive QGs at $z\sim3$ (e.g. \citealt{vanderWel14, Lustig21, Kawinwanichakij25}).

\subsubsection{A dusty star-forming BGG in RO-0959-Sat} \label{sec:BGG_ro0959}
\begin{figure*}
    \centering
    \begin{subfigure}{0.55\linewidth}
        \centering
        \includegraphics[width=\linewidth]{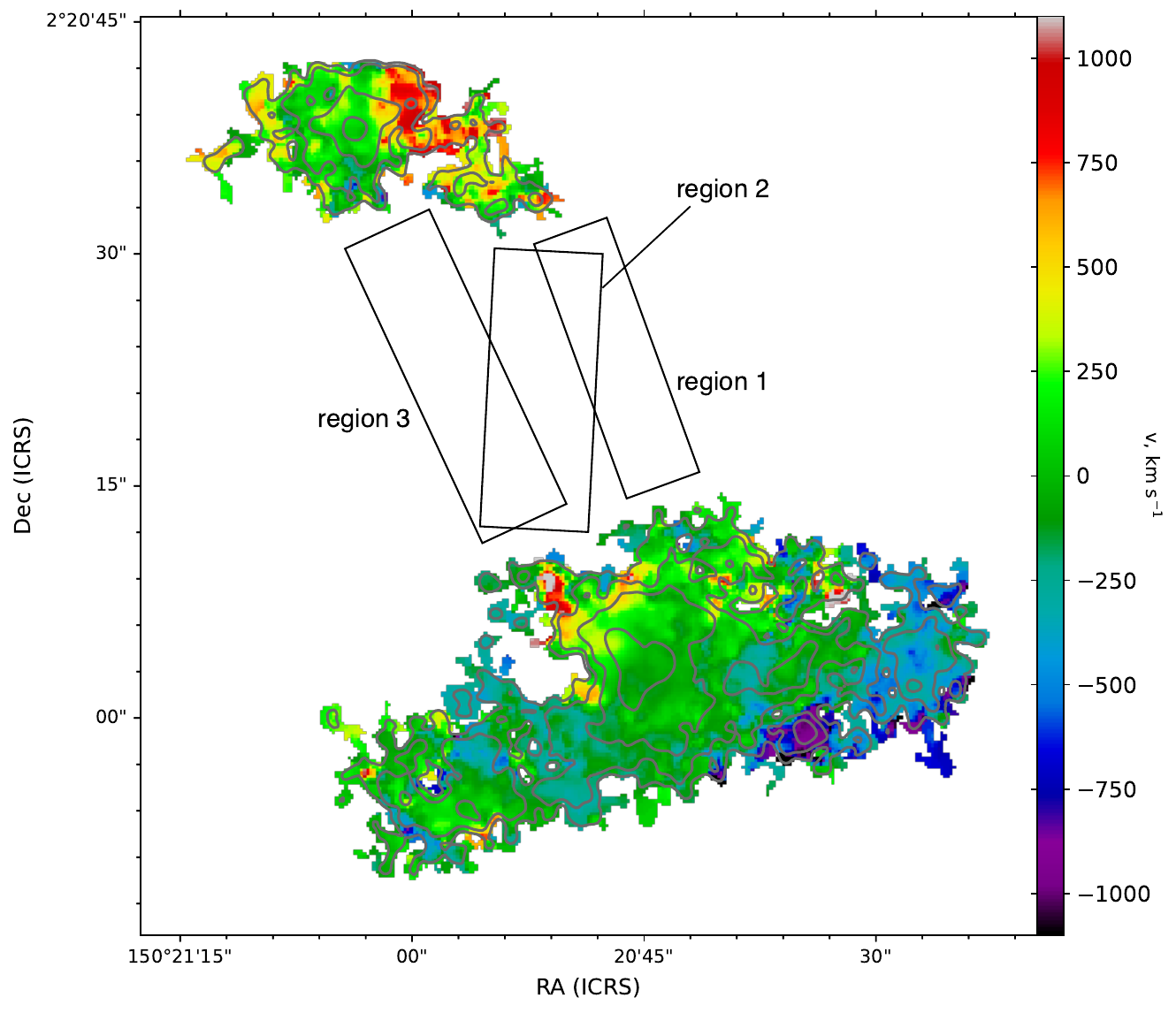}
        \caption{}
        \label{fig:v1_regions_ro1001}
    \end{subfigure}
    \begin{subfigure}{0.42\linewidth}
        \centering
        \includegraphics[width=\linewidth]{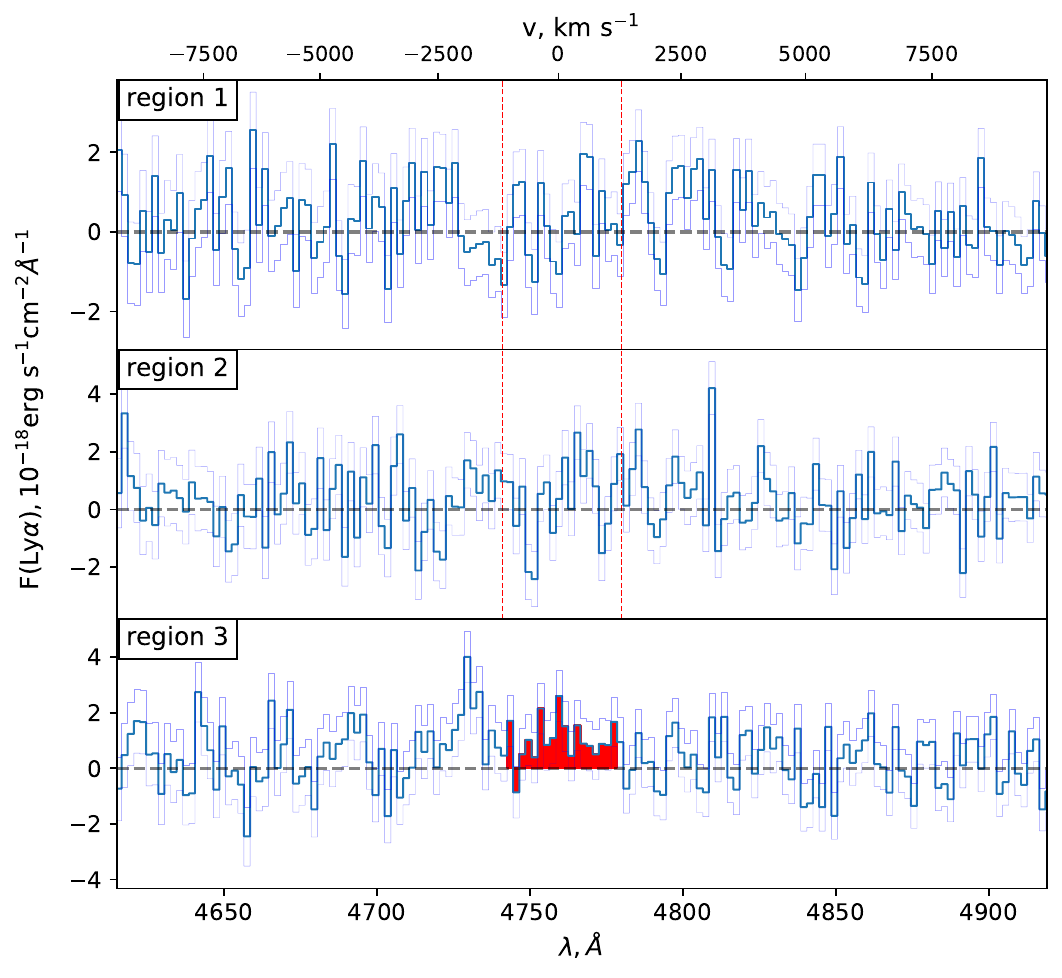}
        \caption{}
        \label{fig:specs_bridges_ro1001}
    \end{subfigure}
    \begin{subfigure}{0.55\linewidth}
        \centering
        \includegraphics[width=\linewidth]{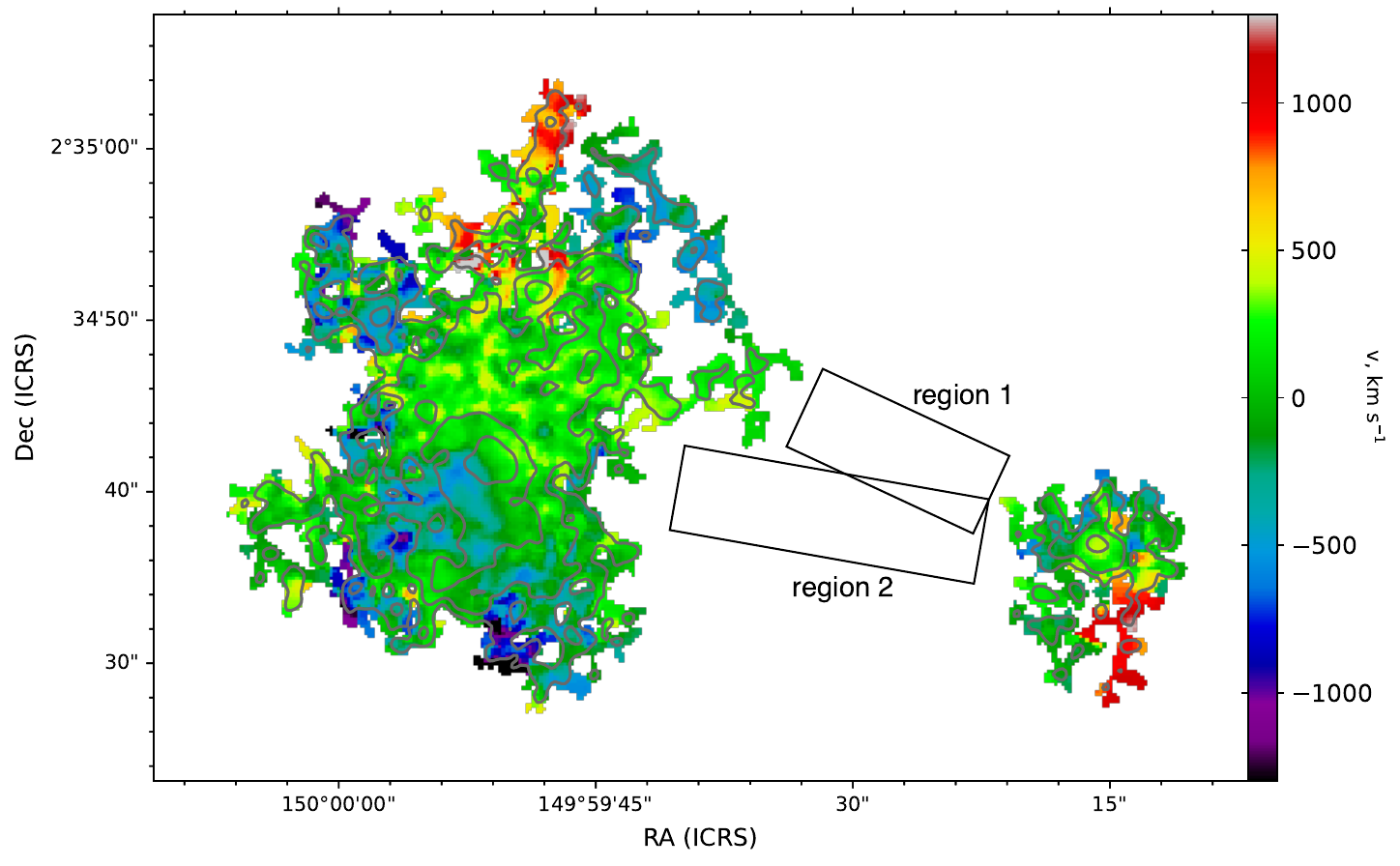}
        \caption{}
        \label{fig:v1_regions_ro0959}
    \end{subfigure}
    \begin{subfigure}{0.42\linewidth}
        \centering
        \includegraphics[width=\linewidth]{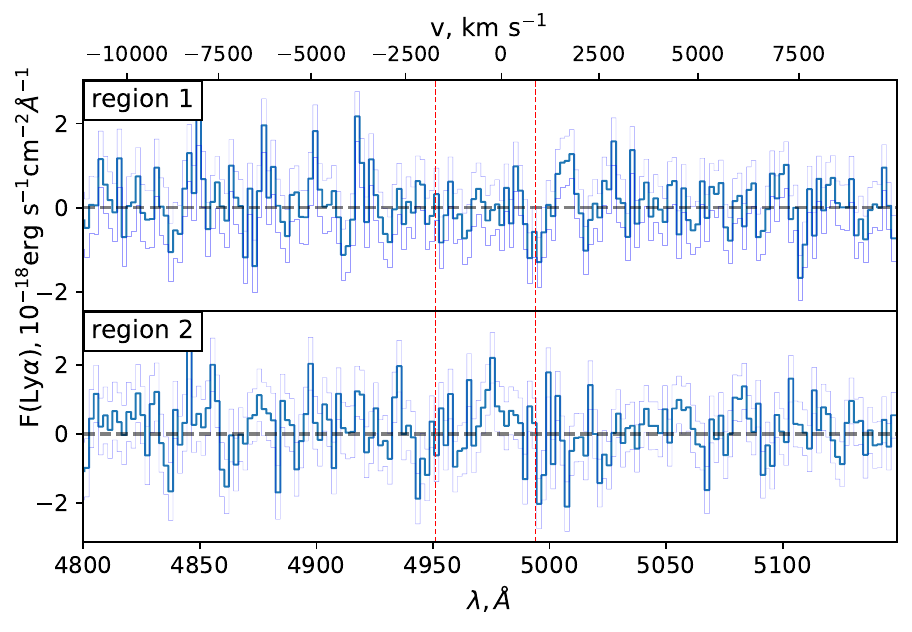}
        \caption{}
        \label{fig:specs_bridges_ro0959}
    \end{subfigure}
    \caption{Velocity maps of (a) RO-1001 and (c) RO-0959 halo pairs. Black boxes illustrate regions where we speculate the possible filaments connecting the two nebulae might be at if there are any. Corresponding integrated spectra extracted from the rectangular regions and re-binned by 2$\AA$ are shown in (b) and (d). Wavelengths marked in red are Ly$\alpha$ emission ranges from the main nebulae. We expect Ly$\alpha$ emission from potential filaments would be located around the same ranges.}
    \label{fig:v1_specs_bridges}
\end{figure*}

The BGG in RO-0959-Sat is well detected in both FIR (S/N$\sim$13) and radio. The FIR-to-radio best-fit SED infers an infrared luminosity of ${\rm log}(L_{\rm IR}/L_{\odot})=12.48\pm0.08$, and thus an SFR$_\mathrm{IR}= 302^{+61}_{-51}~M_\odot$ yr$^{-1}$. Its stellar mass (COSMOS2020 LePhare) is estimated to be log$(M_*/M_\odot) = 10.93\pm0.18$. We notice that there is a mid-IR AGN component present in the best-fit SED, but it does not give a constraining enough estimate of the $L_\mathrm{AGN,~IR}$. This galaxy is detected in both CO[4-3] and strong CO[3-2] (9$\sigma$) by ALMA (Daddi et al., in prep.) suggesting it to be a massive dusty galaxy at $z$ = 3.091, agreeing with the redshift ($z$ = 3.092) of the LAN derived from the spectrum in Fig. \ref{fig:lya_specs}. The expected position of the peak of the Ly$\alpha$ spectrum from this CO(3-2) induced $z$ is marked by the dashed green line in Fig. \ref{fig:lya_specs}, nearly overlapping with the flux weighted Ly$\alpha$ peak, implying that this galaxy indeed resides at the centre of RO-0959-Sat. 

We analysed the morphology of this galaxy following the same method as described in Sect. \ref{sec:BGG_ro1001}. We fitted two Sérsic profiles for the BGG to account for a possible core component and a second offset profile simultaneously for a nearby bright source, alongside a flat sky model. For this galaxy, fitting an additional Sérsic profile improved the fit significantly. Fig. \ref{fig:sersic_fitting} (right) shows the data, the best-fit models, and the residuals. For the disc component, we find a Sérsic index of $n = 0.53\pm0.01$ and an effective radius of $r_{\text{eff}} = 3.77\pm0.03$ kpc, while for the core component, we find $n = 1.36\pm0.10$ and $r_{\text{eff}} = 0.73\pm0.02$ kpc. The low Sérsic indices point to a pseudo-bulge in a disc-dominated galaxy. The Sérsic index of the disc profile is lower than what is typically found for star-forming galaxies (e.g. \citealt{Ward24}). This could be due to the spiral structure seen clearly in the residuals, which flattens the SB profile in the outskirts. The size of the galaxy is in agreement with what has been found in the literature for massive star-forming galaxies at this redshift (e.g. \citealt{vanderWel14, Ward24}).

\subsection{Filament measurements} \label{sec:filaments}
\begin{figure}
    \centering
    \includegraphics[width=\linewidth]{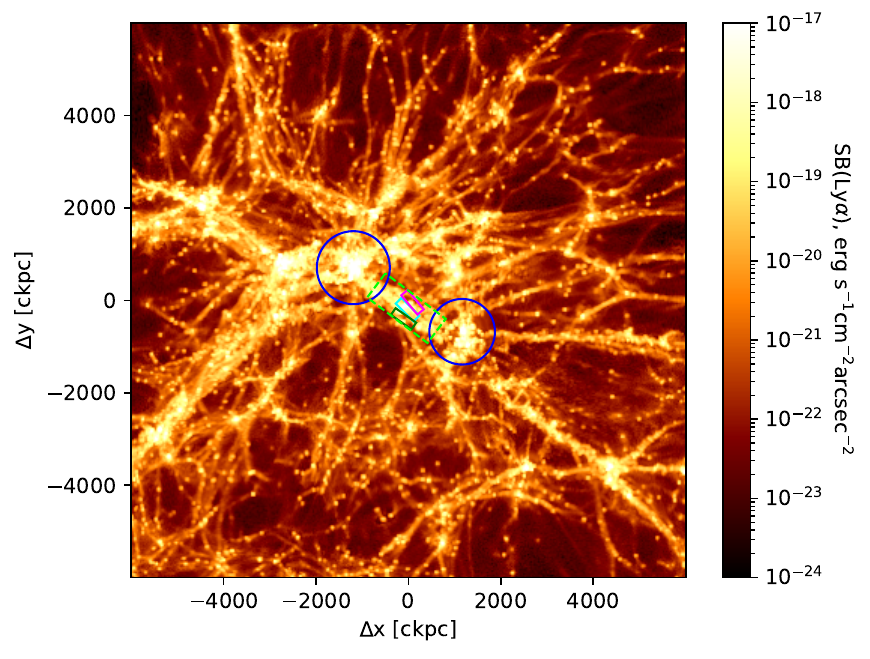}
    \caption{Ly$\alpha$ SB map showing a counterpart pair of halos in the TNG300 simulation projected on the xy plane. The Ly$\alpha$ SB is predicted by the model in \cite{Liu25}. The blue circles show the virial radii of the two halos. The three regions marked by solid rectangles are positioned mimicking regions taken between the two pairs of main and satellite LANs as shown in Fig. \ref{fig:v1_specs_bridges}. The larger region marked by the dashed green rectangle encompasses the whole Ly$\alpha$ filament between the two halos.}
    \label{fig:tng300}
\end{figure}

The small shifts between peaks in Ly$\alpha$ emission lines and the short projected distances between the main and satellite LANs imply that these two halo pairs could be two connected systems. To search for emission from potential filaments bridging the two pairs, we extracted spectra from three regions between RO-1001 and RO-1001-Sat, with sizes of 40$\times$139 kpc$^2$ (region 1), 55$\times$143 kpc$^2$ (region 2) and 48$\times$166 kpc$^2$ (region 3), along with two regions between RO-0959 and RO-0959-Sat being 39$\times$93 kpc$^2$ (region 1) and 39$\times$140 kpc$^2$ (region 2), as is shown in Fig. \ref{fig:v1_specs_bridges}. The choices of the box sizes were made for them to be able to capture as much the filament lengths as possible and at the same time avoid the Ly$\alpha$ emission from either nebula. We placed the boxes in approximate positions between the two halo pairs, where we expect the filaments would possibly reside if there are any. In the field of RO-1001, since we suspected that RO-1001-Sat could be connected with RO-1001 through the seemingly gas-accreting west tail as is discussed in Sect. \ref{sec:satellites}, we placed region 1 between the west tail and the north-west filament in the main nebula, and region 2 between the tail and the velocity peak in the main nebula. Region 3 was placed more randomly in the rest of the blank space between the two LANs. As for the field of RO-0959, we placed region 1 between the filamentary structures in the north of the two LANs, where they appear to be the most close-by in projection, and region 2 between the centres of the two LANs. Following the super-sky correction, noise re-normalisation, and zero-level correction described in Sect. \ref{sec:data_redection}, we have calibrated the overall background level close to a standard Gaussian distribution. We then re-normalised the noise again for each extracted spectrum using the method in Sect. \ref{sec:data_redection} to further improve the noise level. All noise propagations were done quadratically. 

We integrated the spectrum extracted from each box aperture across the wavelength range of Ly$\alpha$ emission in the main nebula to measure the average SB in each region, again with the noise level derived by propagating the variance accordingly and quadratically. In the field of RO-1001, we expected region 1 to be the most likely region to detect signals from filaments, as was stated above. However, we only measure a 3$\sigma$ upper limit on the Ly$\alpha$ SB from this region as SB(Ly$\alpha$) $\lesssim 2.5 \times10^{-19}\mathrm{erg~s}^{-1}\mathrm{cm}^{-2}\mathrm{arcsec}^{-2}$. A similar 3$\sigma$ upper limit is measured from region 2 as SB(Ly$\alpha$) $\lesssim 2.0 \times10^{-19}\mathrm{erg~s}^{-1}\mathrm{cm}^{-2}\mathrm{arcsec}^{-2}$. Notably, in region 3, we have a 5$\sigma$ detection of $4.17 \pm 0.81 \times10^{-19}\mathrm{erg~s}^{-1}\mathrm{cm}^{-2}\mathrm{arcsec}^{-2}$ within the expected range of Ly$\alpha$. Considering that gas in the filaments is likely to have higher velocities than the gas that has already been accreted onto the potential wells of the two halos, if we expand the signal-searching wavelength range to include the adjacent peak to the left of the highlighted range in the spectrum of region 3 (bottom panel in Fig. \ref{fig:specs_bridges_ro1001}), the S/N would increase to 6.9$\sigma$ as $6.42 \pm 0.93 \times10^{-19}\mathrm{erg~s}^{-1}\mathrm{cm}^{-2}\mathrm{arcsec}^{-2}$. However, with a velocity shift of $\sim-1950~\mathrm{km~s}^{-1}$ with respect to the main nebula, if real, this signal is more likely coming from a separate structure, such as a wall-like structure extending further away from the two nebulae. As for RO-0959, region 1 and 2 give 3$\sigma$ upper limits of SB(Ly$\alpha$) $\lesssim (2.8, 2.0)\times10^{-19}\mathrm{erg~s}^{-1}\mathrm{cm}^{-2}\mathrm{arcsec}^{-2}$. To confirm that we were not capturing a higher level of noise distribution in region 3, we checked the sky levels by measuring the S/N distribution in box apertures randomly placed in the background with the average size of the three or two regions placed in each field. Both fields turned out to have sky levels following good Gaussian distributions without significant outliers (Fig. \ref{fig:snr_regions}). Additionally, the average SB levels from these randomly placed apertures are measured to have 3$\sigma$ upper limits within $\lesssim(2.0-2.5)\times10^{-19}\mathrm{erg~s}^{-1}\mathrm{cm}^{-2}\mathrm{arcsec}^{-2}$ and $\lesssim(2.4-6.5)\times10^{-19}\mathrm{erg~s}^{-1}\mathrm{cm}^{-2}\mathrm{arcsec}^{-2}$ in the field of RO-1001 and RO-0959, respectively. We also performed a jackknife resampling in region 3 and obtained a positively skewed distribution of SB not driven by individual pixels (Fig. \ref{fig:jackknife}), implying the presence of extended emission. Given that the noise level does not vary largely across the field, the 5$\sigma$ measurement could indeed hint at some bridging structure between RO-1001 and RO-1001-Sat. However, follow-up observations with longer integration time are needed to confirm this measurement.

To verify these SB measurements of potential filaments, we searched for their counterparts in the TNG300-1 simulation \citep{Pillepich18}, in which the star-forming gas is treated with a sub-grid multi-phase interstellar medium model (\citealt{Springel03}). The Ly$\alpha$ SB in simulated filaments was predicted using the model proposed by \cite{Liu25}, that takes into account the mechanisms of both recombination and collisional excitation, as well as the transmission rates of Ly$\alpha$ photons escaping from dense gas at different redshifts. We find a pair of halos with $M_{\rm h} = (1.7, 1.2) \times 10^{13} M_\odot$ and $R_{\rm vir}$ = (790, 710) ckpc at $z$=3, that are $\sim$2.7 cMpc apart in projection and $\sim$3.7 cMpc apart in 3D, as is shown in Fig. \ref{fig:tng300}. We are not able to find halo pairs that are closer by and connected by a well-defined filament at the same time, likely limited by the box size of the simulation. We placed three boxes with the average size of the five regions (44$\times$136 kpc$^2$) between the two halo pairs in Fig. \ref{fig:v1_specs_bridges} adjacently to mimic the blind searches we did. The average SB in these three boxes are measured to be SB(Ly$\alpha$) = $(0.61, 23.5, 5.2)~\times 10^{-19}\mathrm{erg~s}^{-1}\mathrm{cm}^{-2}\mathrm{arcsec}^{-2}$ from the top to the bottom. The high SB(Ly$\alpha$) in the central region is due to it being accidentally placed on a small halo ($M_\mathrm{h}\gtrsim10^{11}M_\odot$). We also measure the average SB(Ly$\alpha$) of the filament connecting the two halos to be $5.5 \times10^{-19}\mathrm{erg~s}^{-1}\mathrm{cm}^{-2}\mathrm{arcsec}^{-2}$ in the large dashed box encompassing the whole filament in Fig. \ref{fig:tng300}. By and large, these measurements from the simulation are consistent with our observations. However, we are aware that this comparison is based on two single pairs of halos from the observation and simulation, which clearly limits the statistical weight of the result.

\section{Discussion} \label{sec:discussion}
\begin{figure*}
    \centering
    \begin{subfigure}{0.49\linewidth}
        \centering
        \includegraphics[width=\textwidth]{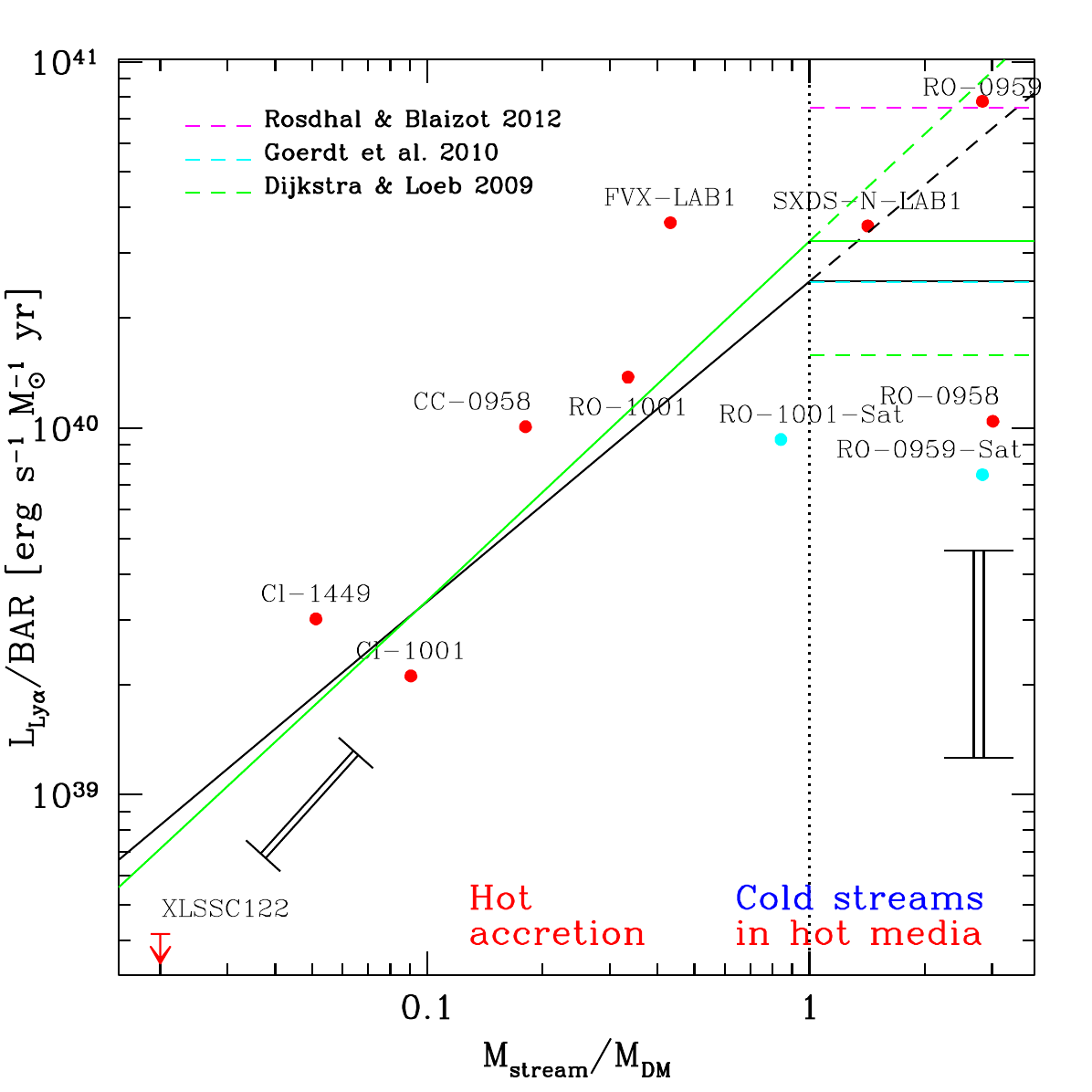}
        \label{fig:llya/bar_mstream/mh}
    \end{subfigure}
    \begin{subfigure}{0.49\linewidth}
        \centering
        \includegraphics[width=\textwidth]{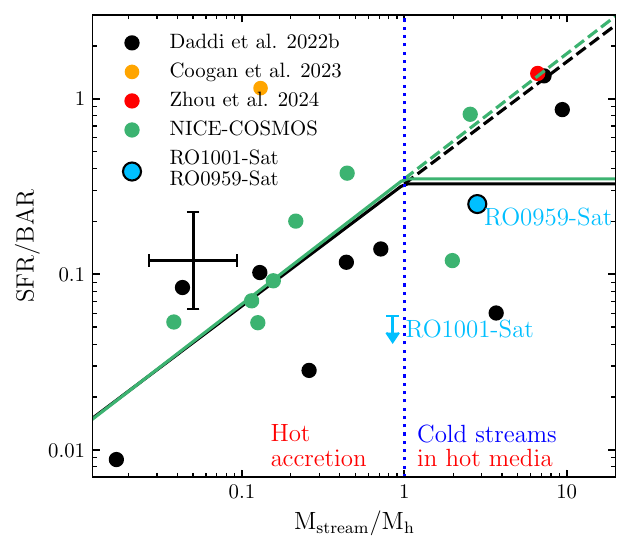}
        \label{fig:sfr/bar_mstream/mh}
    \end{subfigure}
    \caption{\textit{Left}: Revised Fig. 2 from \cite{Daddi22a} showing the distribution of LANs in this work (blue points) and previous KCWI observations (red points) across the plane of $L_\mathrm{Ly\alpha}$/BAR-$M_{\rm stream}/M_{\rm h}$. The green line is the best fit in \cite{Daddi22a} revised using the modified $M_\mathrm{stream}$ in \cite{Daddi22b}. \textit{Right}: Revised Fig. 3 from \cite{Sillassen24} showing how massive galaxy groups in this work (blue points), \cite{Coogan23} (orange points), previous KCWI observations (\citealt{Daddi22a}, black points), and the NICE survey (\citealt{Zhou24} in red points and \citealt{Sillassen24} in green points) are distributed across the SFR$_\mathrm{IR}$/BAR-$M_{\rm stream}/M_{\rm h}$ plane. The green line is the original best fit in \cite{Sillassen24}. In both panels, black lines show the revised best fits of Eq. (4) in \cite{Daddi22a} with $\alpha_\mathrm{Ly\alpha}=0.87\pm0.19$ (\textit{left}) and of Eq. (5) in \cite{Sillassen24} with $\alpha_\mathrm{SFR}=0.70\pm0.15$ (\textit{right}) after adding RO-1001-Sat and RO-0959-Sat into the fits.}
    \label{fig:bar_mstream/mh}
\end{figure*}

\subsection{Clustering of halos} \label{sec:clustering_satellites}
We find that the two pairs of nebulae have remarkably small velocity differences ($\sim$ 100-300 km s$^{-1}$) as shown in Fig. \ref{fig:lya_specs}, compared to the internal velocity dispersion ($\sim$ 500 km s$^{-1}$). This suggests that they might be presumably connected by a single filament of neutral gas, which, if true, is important to set out future merging. It is quite a fundamental expectation that the halo mass function grows rapidly at $z\sim3$ and $M_{\rm h}\sim10^{13}M_\odot$ (e.g. \citealt{Mo02}), which requires massive halos to have nearby halos in their surroundings that we are finding. Being at the same $z$ and close by ($\sim$ 1cMpc projected distance), these two pairs might be the progenitors of local galaxy clusters and will collapse into one by $z\sim0$. 

In principle, the 76\% probability of a satellite halo neighbouring a main halo estimated in Sect. \ref{sec:near_neighbour_prob} should be smaller, since only dark matter halos that contain Ly$\alpha$ emitting gas should be counted. Additionally, distances between some of the selected pairs might be much larger since we only consider the projected distances. However, this should not be a major concern as the majority of the neighbouring halos identified should come from the immediate large-scale structure of the central halos, as chance alignment effects are rare (see Fig. 7 in \citealt{Sillassen24}). Therefore, there is a good agreement between our findings and the simulations. This is certainly based on very small number statistics, and larger surveys on high-$z$ LANs, such as the NOEMA formIng Cluster survEy (NICE; \citealt{Zhou24, Sillassen24}), are needed to search for similar satellite halos for a more robust estimate of the probability of finding such structures.

Nevertheless, it is shown that the occurrence of massive halo pairs should not be rare even at relatively high $z$. They might have been overlooked previously, perhaps due to a lack of good tracers of such structures. \cite{Daddi22a} proposed that the diffuse Ly$\alpha$ emission could be a tracer of massive halos at high $z$. This is reinforced by our findings, where we see that our LANs might reside on overdense nodes in the cosmic web, which could potentially be connected by filaments. Therefore, we suggest that the large-scale Ly$\alpha$ filaments reported in the literature (e.g. \citealt{Umehata19, Tornotti25b}) might encompass collections of individual dark matter halos in which, zoomed in, one could see pairs such as what we have found. We propose that decomposing large-scale overdensities into single dark matter halos, as in a principal component analysis, could be powerful to identify paired or clustered halos.

\subsection{Gas accretion diagrams} \label{sec:bar}
Using Eq. (1) from \cite{Daddi22a} with the estimated $M_\mathrm{h}$ from Sect. \ref{sec:satellites}, we calculate the total BARs of RO-1001-Sat and RO-0959-Sat as 3140$\pm$82 and 1369$\pm$37 $M_\odot$ yr$^{-1}$. The transitional halo mass $M_\mathrm{stream}$ between the `hot accretion' and `cold stream in hot media' regimes is predicted to be $\simeq10^{13.13}$ and $10^{13.25}M_\odot$ at the redshifts of the two LANs following Eq. (2) in \cite{Daddi22b}, placing RO-1001-Sat and RO-0959-Sat separately in the hot and cold accretion regimes. In Fig. \ref{fig:bar_mstream/mh}, we add these two halos to Fig. 2 in \cite{Daddi22a} and Fig. 5 in \cite{Sillassen24}, that illustrate how $L_{\mathrm{Ly}\alpha}$ and SFR$_\mathrm{IR}$ of massive galaxy groups in our MUSE sample, the previous KCWI observations \citep{Daddi22a}, the NICE survey \citep{Zhou24, Sillassen24}, and \cite{Coogan23} are distributed across the $M_{\rm stream}$ boundary between the two regimes. We see that although it falls in the `hot accretion' regime, RO-1001-Sat sits on the border between the two regimes, meaning cold-mode accretion starts to decrease but should still constitute a significant fraction of the gas accreted onto it because the evolution from cold to hot accretion is supposed to be a gradual transition instead of a sharp decline. Both satellite LANs reside below the original best-fit lines, resulting in slightly decreased slopes in the revised fits. Despite having a 0.5 dex higher $L_{\mathrm{Ly}\alpha}$ than RO-0959-Sat, RO-1001-Sat appears to form stars at less than half the rate, highlighting the substantial scatter in the correlation between $L_{\mathrm{Ly}\alpha}$ and SFR, for individual halos.

\subsection{BGG in RO-1001-Sat} \label{sec:BGG_ro1001_discussion}
\subsubsection{Membership to the group} \label{sec:BGG_ro1001_redshift}
There is no emission line detection in the spectrum extracted at the BGG from our MUSE data, except for Ly$\alpha$, which could originate from the nebula instead of the galaxy itself. We tried to fit its spectrum with a set of templates using \texttt{czspecfit}, but could not get any reliable fit due to the low S/N. The $z_{\rm phot}$ of this galaxy from COSMOS2020 is $2.91^{+0.08}_{-0.44}$, close to $z = 2.920$ of RO-1001-Sat calculated from the flux-weighted peak in its Ly$\alpha$ spectrum (Sect. \ref{sec:satellites}). Fig. \ref{fig:sb_v2_ro1001sat} shows this galaxy coincidentally lies on top of the Ly$\alpha$ peak and aligns within uncertainties with the velocity dispersion peak of the nebula (being offset by $\sim0.39\arcsec$ = 3pkpc). Moreover, the peaks of both Ly$\alpha$ SB and velocity dispersion appear to be elongated along the NE-SW axis, mimicking the elliptical shape of the BGG. We estimate the probability of a random LOS alignment of this BGG with the centre of the LAN to be $\sim1.5\times10^{-5}$, by calculating the probability of finding a galaxy with $M_*\geq10^{11.3}M_\odot$ that has the same colour as the BGG or is redder than it (i.e. $J-K\gtrsim J-K_\mathrm{BGG}$), within an aperture of $r=2\arcsec$ in the redshift range used for the member identification ($2.52\lesssim z\lesssim3.31$) from the entire COSMOS field. Furthermore, by accounting for all candidate members in Table \ref{tab:galaxies_ro1001}, we calculate the barycentre of RO-1001-Sat to be at (150.3509737, 2.3438086), only $\sim0.23\arcsec$ (1.8 pkpc) from the BGG (150.351016, 2.343855). Taking all the above into consideration, it is of little doubt that this galaxy indeed resides at the centre of RO-1001-Sat. However, further spectroscopic confirmation is certainly still necessary.

\subsubsection{Quenching mechanisms} \label{sec:BGG_ro1001_QG}
Adopting the $z_\mathrm{phot}$ from COSMOS2020, this BGG is classified as quiescent based on its UVJ colour and optical SED in Sect. \ref{sec:BGG_ro1001}. However, Fig. \ref{fig:sb_v2_ro1001sat} shows that this galaxy is on top of the Ly$\alpha$ SB peak; that is, it sits at the centre of a cold gas nebula. Besides, as stated above, this BGG resides at the centre of the potential well of the host halo that the gas is accreting onto. Moreover, RO-1001-Sat is supposed to reside close to the border between the `cold stream in hot media' and the `hot accretion' regimes illustrated in Fig. \ref{fig:bar_mstream/mh}, where cold accretion should still play a major role, as is discussed in Sect. \ref{sec:bar}. Similar cases of massive QGs in overdense environments have been reported previously in \cite{Kalita21} and \cite{Kubo21}, where the QGs are, however, not at the centre of the LANs. This is the first time a QG is found to be located at the centre of a LAN while still remaining quenched. \cite{Carnall20} also report two early-type galaxies (ETGs) with weak Ly$\alpha$ emission at $z\sim3.4$. Given our observations, we suggest that this Ly$\alpha$ emission line could be arising from surrounding weak nebulae hosting the ETGs rather than the galaxies themselves.

A similar QG, ADF22-QG1, reported by \cite{Kubo21} also resides in a LAN with comparable scales as RO-1001-Sat, but it seems to sit on top of a weak bridge between two Ly$\alpha$ peaks \citep{Umehata19}. A molecular gas reservoir in this galaxy detected in CO(3-2) is reported recently by \cite{Umehata25}. The QG, galaxy-D, identified in the main nebula of RO-1001 by \cite{Kalita21}, is the closest resemblance to our BGG since they both reside in a group environment where cold gas is accreting onto, and at the same time have comparable ages (0.1 dex difference). Only $z_\mathrm{phot}$ of galaxy-D is reported in \cite{Kalita21}, but its redshift has been confirmed spectroscopically by Keck later (private communication). However, galaxy-D is not at the centre of the potential well or the peak of Ly$\alpha$ emission, as the BGG is.

Fig. \ref{fig:uvj_sed_cosmos} shows that the BGG is the only QG among all candidate group members in RO-1001-Sat, which is a reversed scenario compared to the typical environmental quenching, where central galaxies tend to be star-forming surrounded by quiescent satellite galaxies in galaxy groups and clusters. Alternatively, this BGG could have quenched its star formation while still retaining a relatively high cold gas fraction in its disc, prevented from further collapse by its stellar bulge through morphological quenching (e.g. \citealt{Martig09}). However, the Sérsic profile fitting does not reveal an evident bulge component. Deeper ALMA observations will be needed to constrain the gas fraction in the BGG more stringently.

A more feasible scenario could be due to the radio AGN feedback. We see a radio excess at RO-1001-Sat as well as the discrepancy between the SFRs estimated from the optical SED fitting ($2.51^{+0.50}_{-0.06}~M_\odot$ yr$^{-1}$) and $L_{\rm radio}$ (300 $M_\odot$ yr$^{-1}$) in Sect. \ref{sec:BGG_ro1001}. For the SFR derived from $L_{\rm radio}$ to be true, there needs to be hidden dust-extinguished star formation, which either is disfavoured by the non-detection in ALMA or requires deeper observations to be revealed. Alternatively, the presence of a radio AGN could explain the overestimation of SFR from $L_{\rm radio}$ compared to the optical SED. And the BGG could thus be quenched and kept quiescent by the cycle of a radio-mode AGN, i.e. the AGN expels gas out of the galaxy, dims and reactivates after the galaxy replenishes its gas supply, similar to the radio AGNs with high duty cycles found in the datasets of \cite{D'Eugenio20,D'Eugenio21} and a sample of quiescent radio galaxies at $z\sim1-5$ with very luminous radio AGNs ($>10^{26}\mathrm{W~Hz}^{-1}$) reported in \cite{Falkendal19}. \cite{Chang24} find that the radio-mode feedback does not affect the cold gas properties on large scales, which is consistent with us seeing an enhancement in Ly$\alpha$ emission and velocity dispersion at the centre, overlapping with the galaxy. Since enhanced velocity dispersion is usually perpendicular to the radio jet \citep{Couto&Storchi-Bergmann23}, the local enhancement of the velocity dispersion around the BGG could be due to a mild radio jet inclined at some angle to the plane of sight. However, due to the low resolution of the radio data, we cannot see any excess of radio jet(s). Future deeper radio observations will be needed to confirm the presence of a radio AGN. 

\subsubsection{Ly$\alpha$ powering source} \label{sec:BGG_ro1001_lya_powering}
Due to the BGG being quenched, the LAN cannot be powered by ionising photons from star formation in the galaxy, and hence alternative powering mechanism(s) are needed. Since the radio-to-FIR SED yields a negligible AGN component, if an AGN is in fact present in the halo, it would either have a dusty torus that is highly obscured in the optical or just a very low bolometric luminosity. If the nebula is powered by the AGN photoionisation, using the AGN bolometric luminosity required to power RO-1001 from \cite{Daddi21} and scaling it down with respect to $L_\mathrm{Ly\alpha}$ of the two nebulae, an AGN would need to have $L_\mathrm{AGN,bol}\sim 10^{44.9}~\mathrm{erg~s}^{-1}$ in order to produce enough ionising photons. The non-detection in X-ray gives us a 3$\sigma$ upper limit of $L_\mathrm{AGN,bol}\lesssim 10^{44.9}~\mathrm{erg~s}^{-1}$ for the BGG, consistent with the minimum required $L_\mathrm{AGN}$. However, considering the typically low Lyman continuum escape rate ($\sim$30\%) for AGN ionising photons, even if there is a weak AGN present in the halo, it would not be powerful enough to ionise the entire LAN. Another possible mechanism is through collisional excitation by shocks (from e.g. radio jets or gas accretion). Scaling down the gravitational energy associated with the cosmological gas accretion available in RO-1001 \citep{Daddi21}, $M_\mathrm{h}$ of RO-1001-Sat gives gravitational energy $\sim100\times$ what is required to power the Ly$\alpha$ emission. Thus, RO-1001-Sat likely shares a similar powering mechanism for its LAN as RO-1001.

\subsection{BGG in RO-0959-Sat} \label{sec:BGG_ro0959_discussion}
RO-0959-Sat has far fewer candidate member galaxies identified and is in a less dense environment than RO-1001-Sat, and hence the central galaxy is supposed to have less matter to accrete. Interestingly, this central galaxy has still been able to form a similar stellar mass, $M_* \approx 10^{11.2}M_\odot$, as the BGG in RO-1001-Sat ($\approx10^{11.3}M_\odot$). This could have been attained through small mergers of nearby galaxies, which might explain why there is nearly no candidate member galaxy around. This merging activity would have likely created a pseudo-bulge at the centre of the galaxy as revealed by the Sérsic profile fitting in Sect. \ref{sec:BGG_ro0959}. The CO(3-2) detection indicates that there is available gas in the interstellar medium to sustain star formation. On the other hand, the non-detection in either radio or X-ray indicates that there is no bright AGN present ($L_\mathrm{AGN,bol}\lesssim 10^{45.1}~\mathrm{erg~s}^{-1}$), which suggests that there is no strong feedback pushing gas outside of the galaxy, and thus the accreted gas was able to further cool down to fuel the star formation, opposite to the case of the BGG in RO-1001-Sat as discussed in Sect. \ref{sec:BGG_ro1001_QG}. The star formation could also be the source of ionising photons powering the LAN. 

The findings of two central galaxies with distinctively different morphologies in the two similar satellite LANs at similar $z$ are puzzling. One explanation could be that as RO-1001-Sat is more massive, the system is more mature and has assembled more satellite galaxies. The BGG in RO-1001-Sat is likely more evolved than in RO-0959-Sat (assuming a downsizing scenario), which is evidenced by it being quiescent. Moreover, the system clearly shows a more mature structure resembling a galaxy cluster with a BGG clearly distinguished from other member galaxies in mass and brightness. Hypothetically, if RO-1001-sat was observed at an earlier stage of evolution, it would possibly be similar to RO-0959-Sat. Conversely, RO-0959-Sat could be the more evolved system where the central galaxy has merged almost all the group members as discussed above, and the RO-1001-Sat BGG is also in this process of mass assembly. Follow-up observations through NIRSpec IFS (with PRISM) would enable full-spectral resolved fitting of each galaxy, allowing for a much more refined SFH determination, which is essential to unveil what caused the divergence.

\subsection{Cosmic filament detection} \label{sec:filaments_discussion}
We measured the SB(Ly$\alpha$) in regions with an average size of $\sim(0.18\times0.54)\, \rm{cMpc}^2$ at different possible locations connecting the two pairs of nebulae in an attempt to unveil signals from bridges in between the nebulae, if there are any. The best measurement we obtain is $(4.17 \pm 0.81) \times10^{-19}~\rm{erg~s}^{-1}{cm}^{-2}{arcsec}^{-2}$ in the field of RO-1001, consistent with the prediction from the TNG300 simulation ($5.5 \times10^{-19}~\rm{erg~s}^{-1}{cm}^{-2}{arcsec}^{-2}$). We see that the predicted SB(Ly$\alpha$) varies with location, which also agrees with the measurements from our blind searches. However, a direct comparison between the two values should be made with caution, as multiple factors could complicate the predicted SB levels in simulated filaments. As the filament selected spans $\sim$1.6 cMpc, for it to have the same level of SB(Ly$\alpha$) as the $\sim$0.54 cMpc regions that we put between the two pairs, shorter filaments in the simulation could appear brighter. On the other hand, as this filament overlaps with a small halo as mentioned in Sect. \ref{sec:filaments}, that enhances its SB level, shorter filaments might not necessarily be brighter if no such halo is present. Previous studies have found two Ly$\alpha$ filaments connecting galaxies in the SSA 22 protocluster at $z\sim3.1$ \citep{Umehata19}, where they measured the SB of the extended Ly$\alpha$ emission to be $\gtrsim 3\times10^{-19}\rm{erg s}^{-1}\rm{cm}^{-2}\rm{arcsec}^{-2}$, consistent with our measurements. Even fainter filaments have also been revealed by several studies. \cite{Bacon21} report extended Ly$\alpha$ emission around LAE overdensities from $z\sim3.1-4.5$ at SB levels of $3-8\times10^{-20}~\rm{erg~s}^{-1}{cm}^{-2}{arcsec}^{-2}$. \cite{Tornotti25a} discover a cosmic web filament in Ly$\alpha$ connecting two QSOs at $z\sim3.22$ with a mean SB$_{\rm{Ly}\alpha} = (8.1\pm0.6)\times10^{-20}~\rm{erg~s}^{-1}{cm}^{-2}{arcsec}^{-2}$, and \cite{Tornotti25b} detect another $\sim$ 5 cMpc portion of a cosmic web filament at $z\sim4$ in Ly$\alpha$ emission within an overdensity of 19 LAEs with a similar SB$_{\rm{Ly}\alpha}$ of the diffuse emission, $3\times10^{-20}~\rm{erg~s}^{-1}{cm}^{-2}{arcsec}^{-2}$. The lower SB$_{\rm{Ly}\alpha}$ in the three cases could be due to the lower $M_{\rm h}$ of the hosting halos, and similarly for the filaments. 

Fig. \ref{fig:lya_specs} shows that the Ly$\alpha$ spectra from both satellite halos are blueshifted, hinting at gas infall. Moreover, both satellites are expected to reside in the cold stream regime as illustrated in Fig. \ref{fig:bar_mstream/mh}. If they are connected to the main halos by filament(s), such filament(s) could be the funnel(s) of gas accretion onto the satellite halos. With deep enough observations, we hope to reveal the filament properties such as transition radii and gas kinematics. For now, we have a $5\sigma$ tentative measurement for a potential bridge connecting the main and satellite nebulae of RO-1001 after 8h of integration time. We will need $\gtrsim$ 32h to double the S/N of the integrated SB of the tentative filament. For this potential filament to be visible in the narrow-band image, it would require S/N $\gtrsim$ 3 in each smoothed pixel ($\sim1''\times1''$), which would need $\gtrsim36$ times the current on-source integration time based on a crude estimation. This would reach even lower SB limits than the current deepest MUSE fields, MXDF and MUDF (e.g. \citealt{Bacon21, Tornotti25b}), and thus would not be practicable.

\section{Conclusions} \label{sec:conclusions}
We present MUSE follow-up observations of two galaxy groups, RO-1001 and RO-0959, at $z\sim3$, where giant LANs have been found previously by KCWI. We discover two satellite halos at $\sim$1 cMpc from the main halos, and we estimate their hosting halo masses to be log$(M_{\rm h}, M_\odot) = 13.2\pm0.3$ and 12.8 $\pm$ 0.3 for RO-1001-Sat and RO-0959-Sat, respectively. Our findings can be summarised below:
\begin{itemize}
    \item The Ly$\alpha$ spectra extracted from the two pairs of satellite and main LANs show very small velocity shifts of $\sim$ 100-300 km s$^{-1}$, suggesting that within each pair the two systems might be connected by some gas filaments. Our findings reinforce the idea that LANs may be used as tracers of massive dark matter halos.
    \item We estimate from simulations that the probability of finding a satellite halo with $10^{12.5}-10^{13}M_\odot$ within a 1.5 cMpc projected distance from the main halo of $10^{13}M_\odot$ is $\simeq$ 76\%, which is consistent with our observations.
    \item The central brightest galaxy in RO-1001-Sat is quiescent, while the one in RO-0959-Sat is star-forming. This BGG in RO-1001-Sat is the first QG to be observed to date that resides at the centre of both a cold gas nebula and the potential well of an overdensity, making a unique case study for quenching mechanisms in high $z$ overdense environments. We suggest that its quenching mechanism is likely to be radio-mode AGN feedback.
    \item We find evidence of the presence of a cold gas filament between the halo pair of RO-1001 using an integrated spectrum, while the lower-mass system RO-0959 does not present such a signature, consistent with predictions from simulations. 
\end{itemize}

Our findings demonstrate the power of MUSE and future BlueMUSE (e.g. \citealt{Richard19}) to map the large-scale structure traced by extended Ly$\alpha$ emission at high $z$. Large surveys to search for satellite structures of protoclusters in the future will help us clear up the mystery around the assembly history of massive galaxy clusters. At the same time, ALMA observations of molecular gas tracers, such as [CII] and CO(3-2), in this massive QG at the centre of RO-1001-Sat will enable us to connect quenching to gas reservoirs present in the large-scale structure in the high-$z$ Universe.

\begin{acknowledgements}
We thank the referee for constructive comments. We thank Tanya Urrutia for discussions. We are grateful to Yizhou Liu for sharing their modelling results with us and for followup discussions. MA is supported by FONDECYT grant number 1252054, and gratefully acknowledges support from ANID Basal Project FB210003 and ANID MILENIO NCN2024\_112. ID acknowledges funding by the European Union – NextGenerationEU, RRF M4C2 1.1, Project 2022JZJBHM: "AGN-sCAN: zooming-in on the AGN-galaxy connection since the cosmic noon" - CUP C53D23001120006. SG is supported by the LabEx UnivEarthS, ANR- 10-LABX-0023 and ANR-18-IDEX-0001. SJ acknowledges the Villum Fonden research grants 37440 and 13160. M. S. acknowledges support from ANID BASAL project FB210003. This paper makes use of the following ALMA data: ADS/JAO.ALMA\#2013.1.00884.S, ADS/JAO.ALMA\#2016.1.01208.S. ALMA is a partnership of ESO (representing its member states), NSF (USA) and NINS (Japan), together with NRC (Canada), NSTC and ASIAA (Taiwan), and KASI (Republic of Korea), in cooperation with the Republic of Chile. The Joint ALMA Observatory is operated by ESO, AUI/NRAO and NAOJ. This work is based in part on observations made with the NASA/ESA/CSA James Webb Space Telescope. The data were obtained from the Mikulski Archive for Space Telescopes at the Space Telescope Science Institute, which is operated by the Association of Universities for Research in Astronomy, Inc., under NASA contract NAS 5-03127 for JWST. These observations are associated with programme \#1727. The authors acknowledge the COSMOS-Web team for developing their observing programme with a zero-exclusive-access period. The data described here may be obtained from \url{https://dx.doi.org/10.26131/IRSA178}.
\end{acknowledgements}

\bibliographystyle{aa}

\begin{appendix} \label{appendix}
\section{Signal-to-noise ratio maps} \label{app:snr_maps}
\begin{figure}[h]
    \centering
    \begin{subfigure}{0.5\textwidth}
        \centering
        \includegraphics[width=\textwidth]{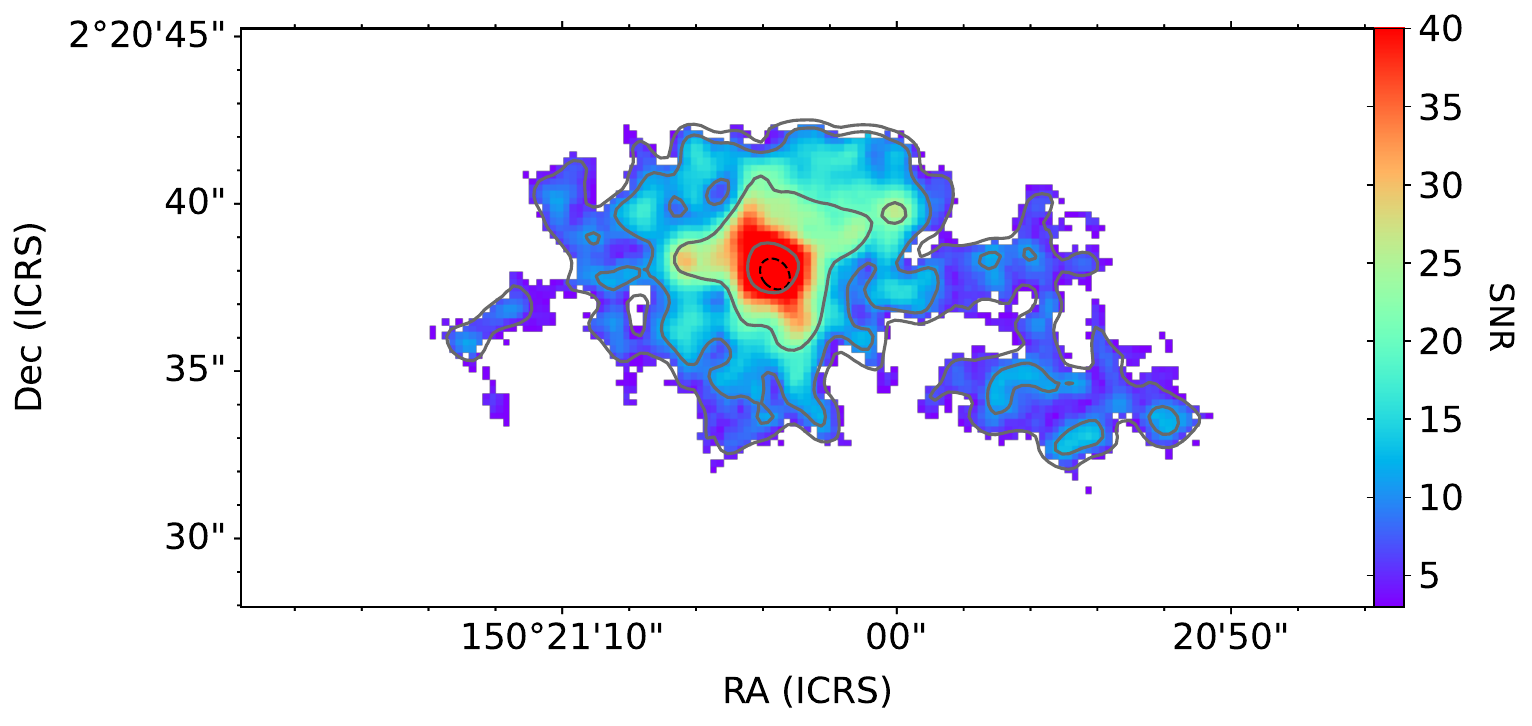}
    \end{subfigure}
    \begin{subfigure}{0.35\textwidth}
        \centering
        \includegraphics[width=\textwidth]{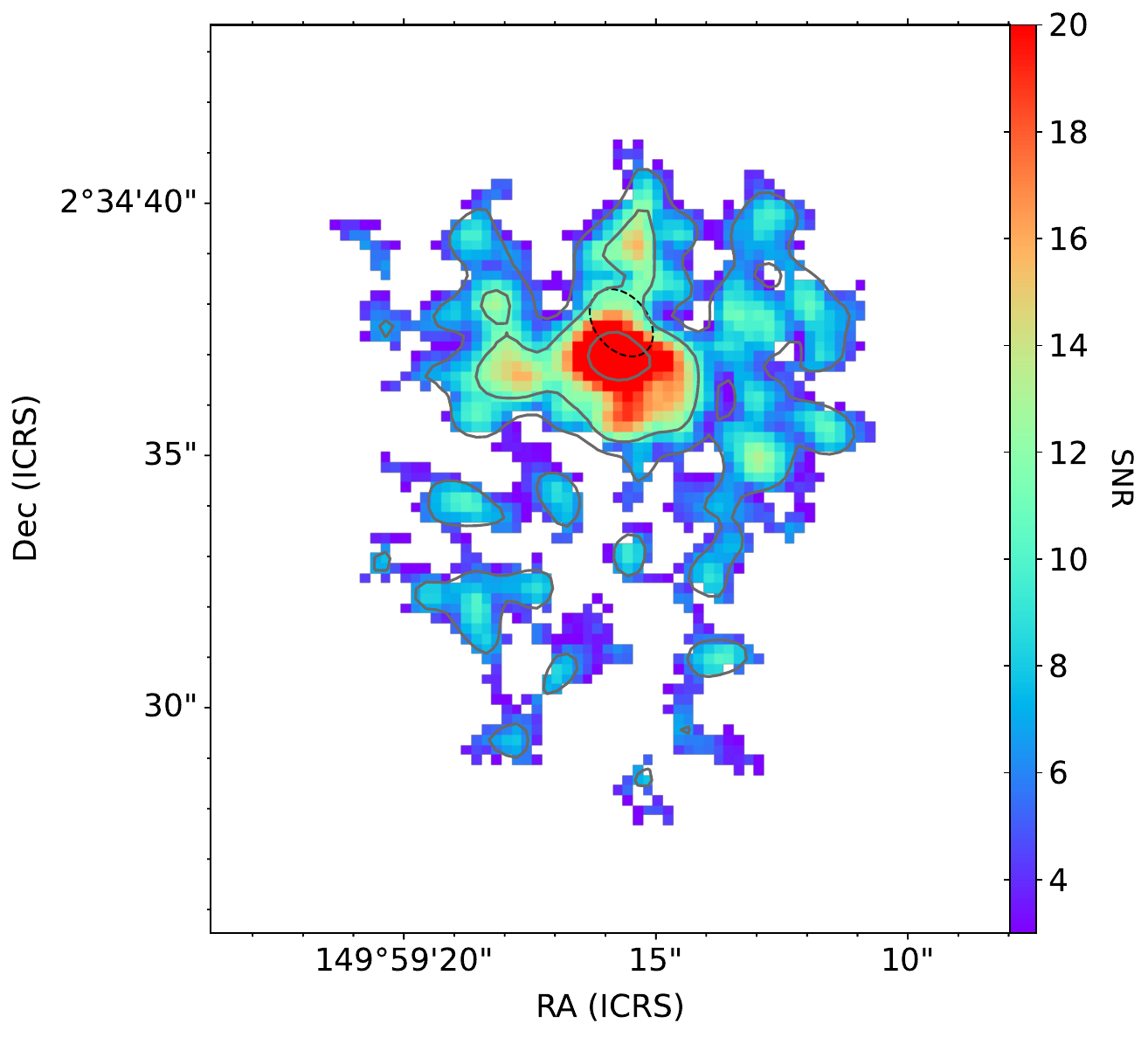}
    \end{subfigure}
    \caption{S/N maps of Ly$\alpha$ SB in RO-1001-Sat (\textit{left}) and RO-0959-Sat (\textit{right}) from AKS-smoothed data. The contours show the same SB(Ly$\alpha$) levels as in Fig. \ref{fig:rgb_ro1001} and \ref{fig:rgb_ro0959}.}
    \label{fig:snrs}
\end{figure}

\section{MultiDark simulations} \label{app:multidark}
\begin{figure}[h]
    \centering
    \includegraphics[width=0.5\textwidth]{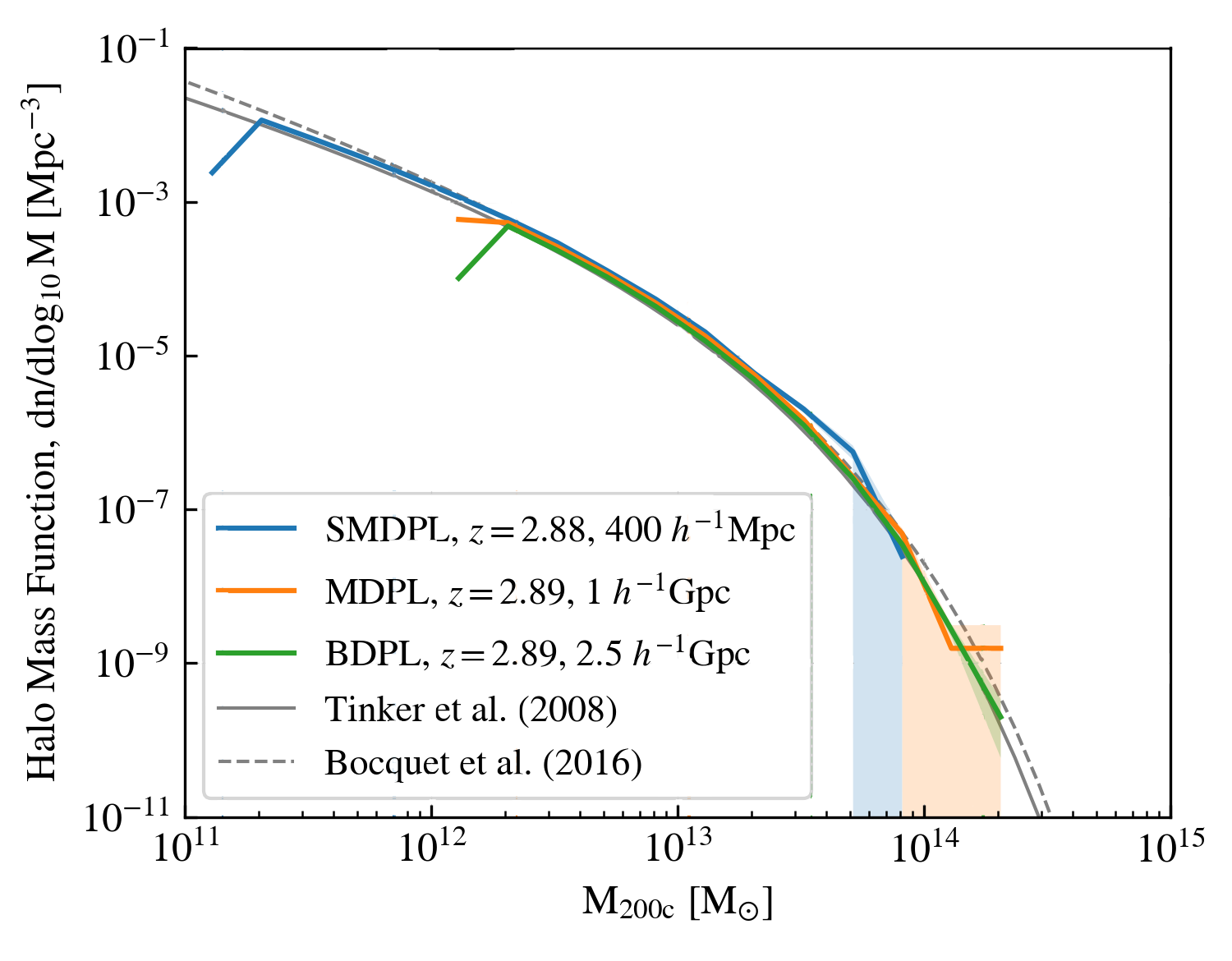}
    \caption{Halo mass functions of the three simulations, SMDPL (green), MDPL (orange) and BMDPL (purple) at $z=2.9$. The predicted halo mass functions by \citet{Tinker08} and \citet{Bocquet16} are plotted in solid and dashed grey lines, respectively. Among the two predictions, \citet{Tinker08} agree best with the simulations, while \citet{Bocquet16} overestimate halo abundance below $10^{11}\rm{M}_{\odot}$ and above $10^{14}~\rm{M}_{\odot}$.}
    \label{Fig:mass_func}
\end{figure}

To avoid being heavily affected by the cosmic variance, we needed to use a sufficiently large cosmological box that includes the expected number of massive halos at $z\sim3$ as indicated by the halo mass function. We chose to use the MultiDark simulation suite, that is a set of three dark matter-only cosmological simulations including SMDPL, MDPL, and BMDPL, each containing $3840^{3}$ particles, with a particle mass resolution of $9.6\times 10^{7}$, $1.9\times 10^{9}$, and $2.4\times 10^{10}$ $h^{-1}\rm{M}_{\odot}$, respectively. These simulations were run using the L-GADGET-2 code (\citealt{Springel05}), optimised for large particle counts, and the Adaptive Refinement Tree (ART) code (\citealt{Kravtsov97,Gottloeber08}). We analysed the simulations using halo finders BDM (\citealt{Klypin97}), RockStar (\citealt{Behroozi13}), and FOF (\citealt{Riebe13}). Halo catalogues were obtained from the CosmoSim archive\footnote{\url{https://www.cosmosim.org/cms/data/projects/multidark-bolshoi-project/}}. Fig. \ref{Fig:mass_func} shows the halo mass functions at redshifts $z\simeq2.9$ for the three simulations compared with the ones predicted by \citet{Tinker08} and \citet{Bocquet16}. It demonstrates excellent convergence across simulations and helps us identify the mass ranges where each simulation provides reliable halo abundances, being log$(M_\mathrm{h}/M_\odot)=$(11-13), (12-14), and (12-14) for SMDPL, MDPL, and BMDPL, respectively. Below these limits, the resolution is insufficient to resolve halos with fewer than 100 particles, while above these limits, the simulation volumes are too small to model large-scale cosmic variance.

\section{Sky level checks and jackknife resampling} \label{app:noise_checks}
\begin{figure*}
    \centering
    \begin{subfigure}{0.43\linewidth}
        \centering
        \includegraphics[width=\linewidth]{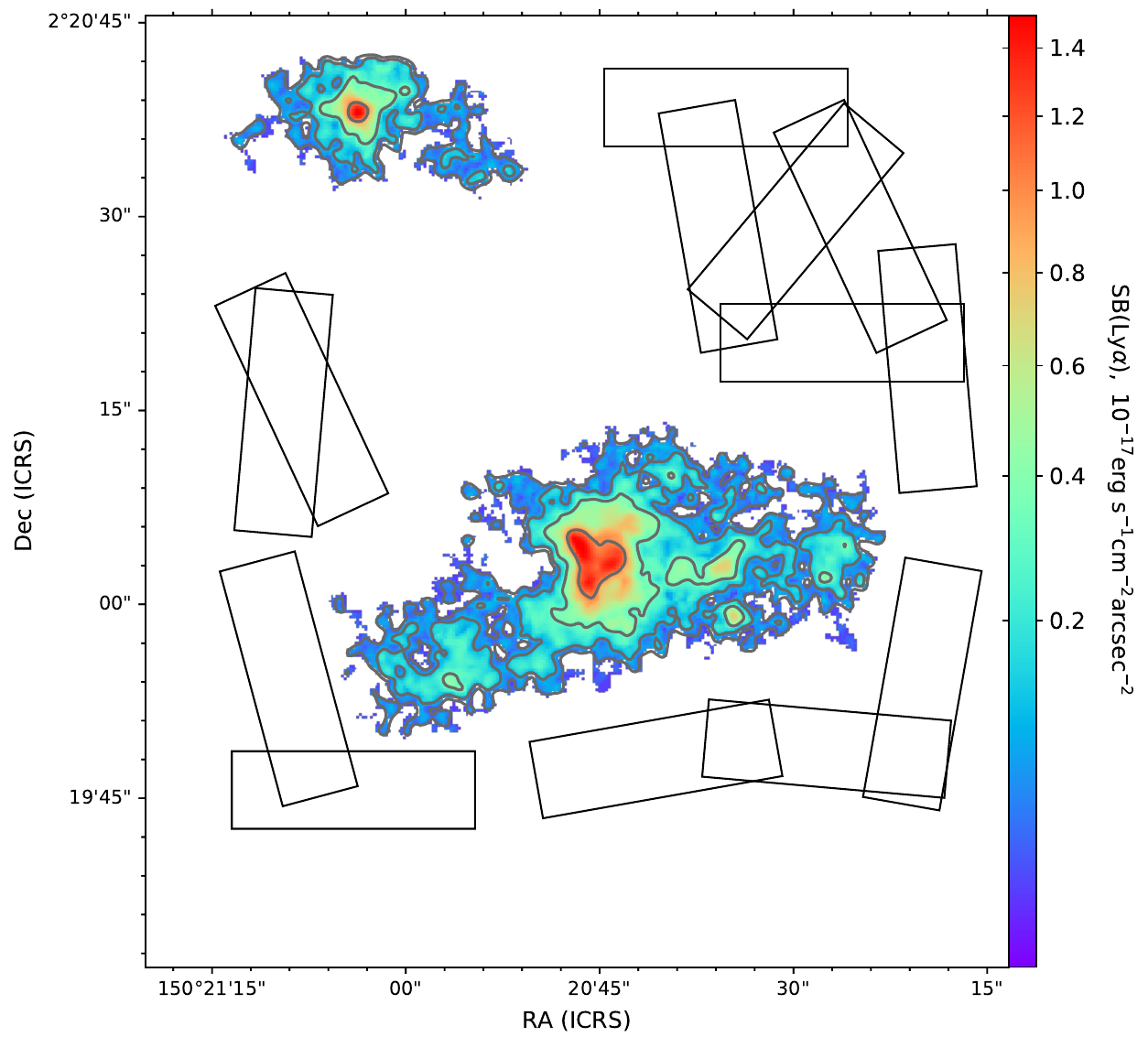}
        \caption{}
        \label{fig:noise_regions_ro1001}
    \end{subfigure}
    \begin{subfigure}{0.43\linewidth}
        \centering
        \includegraphics[width=\linewidth]{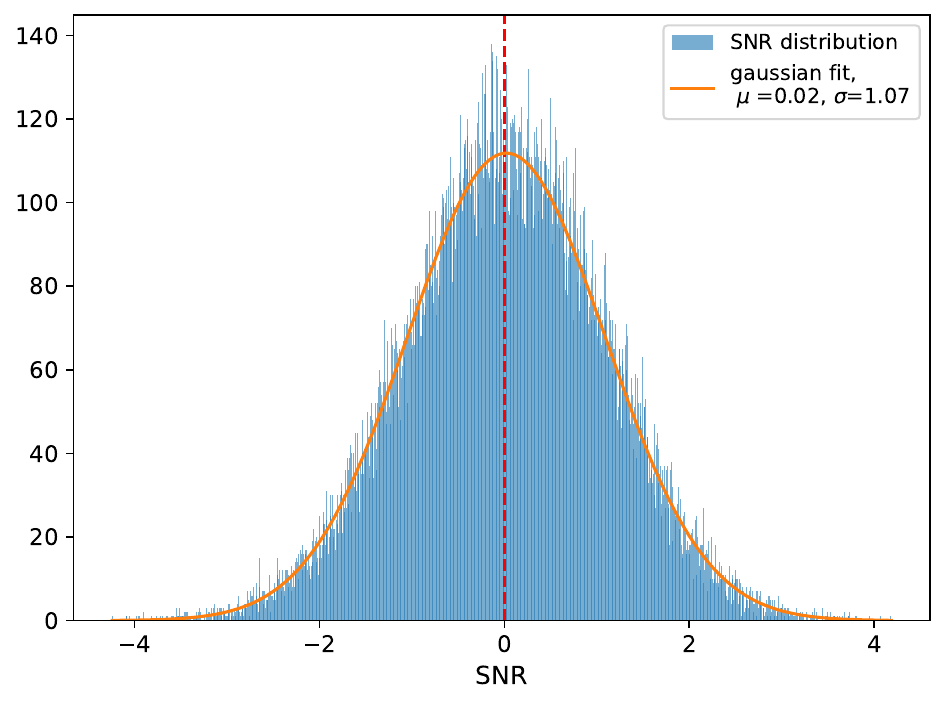}
        \caption{}
        \label{fig:snr_hist_ro1001}
    \end{subfigure}
    \begin{subfigure}{0.43\linewidth}
        \centering
        \includegraphics[width=\linewidth]{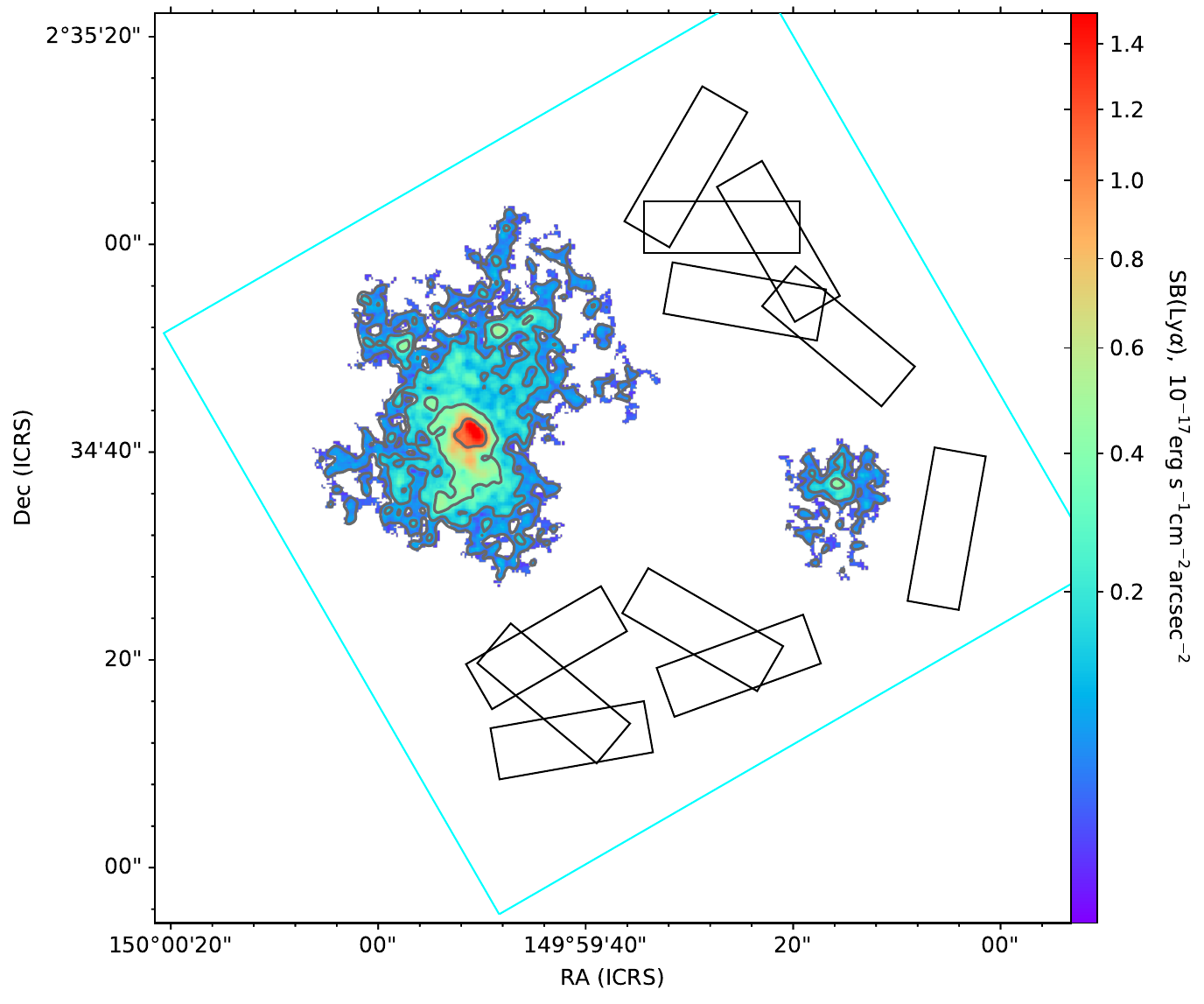}
        \caption{}
        \label{fig:noise_regions_ro0959}
    \end{subfigure}
    \begin{subfigure}{0.43\linewidth}
        \centering
        \includegraphics[width=\linewidth]{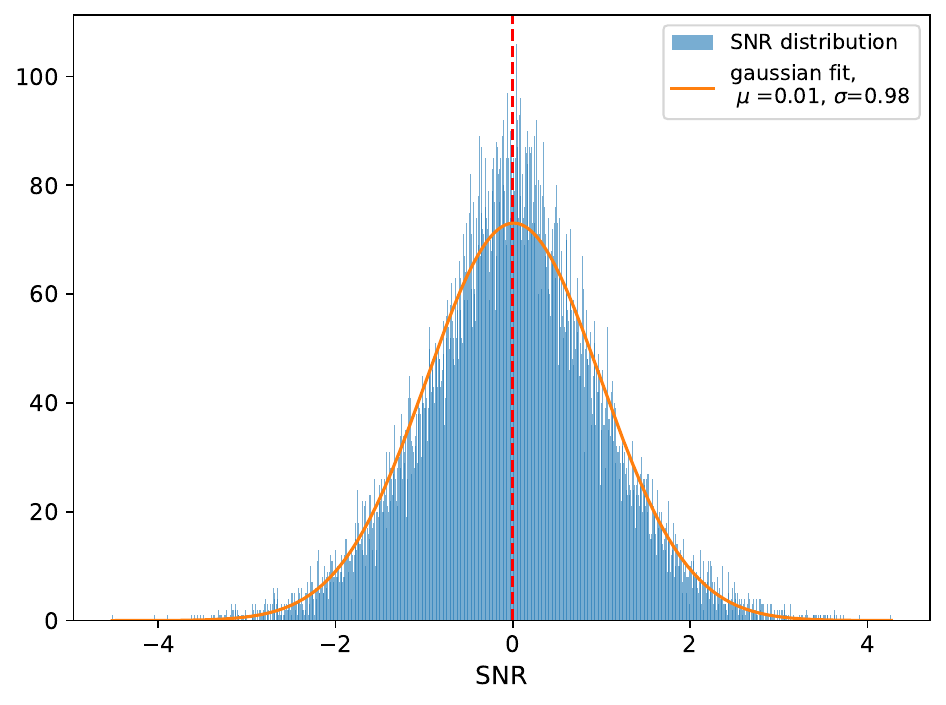}
        \caption{}
        \label{fig:snr_hist_ro0959}
    \end{subfigure}
    \caption{Ly$\alpha$ SB maps of (a) RO-1001 and (c) RO-0959 with black boxes showing apertures selected for the sky level check. The cyan box in (c) shows the MUSE FOV. (b)\&(d) the S/N distributions of pixels in the apertures in the left panels fit with a Gaussian distribution.}
    \label{fig:snr_regions}
\end{figure*}
\begin{figure*}
    \centering
    \includegraphics[width=0.5\textwidth]{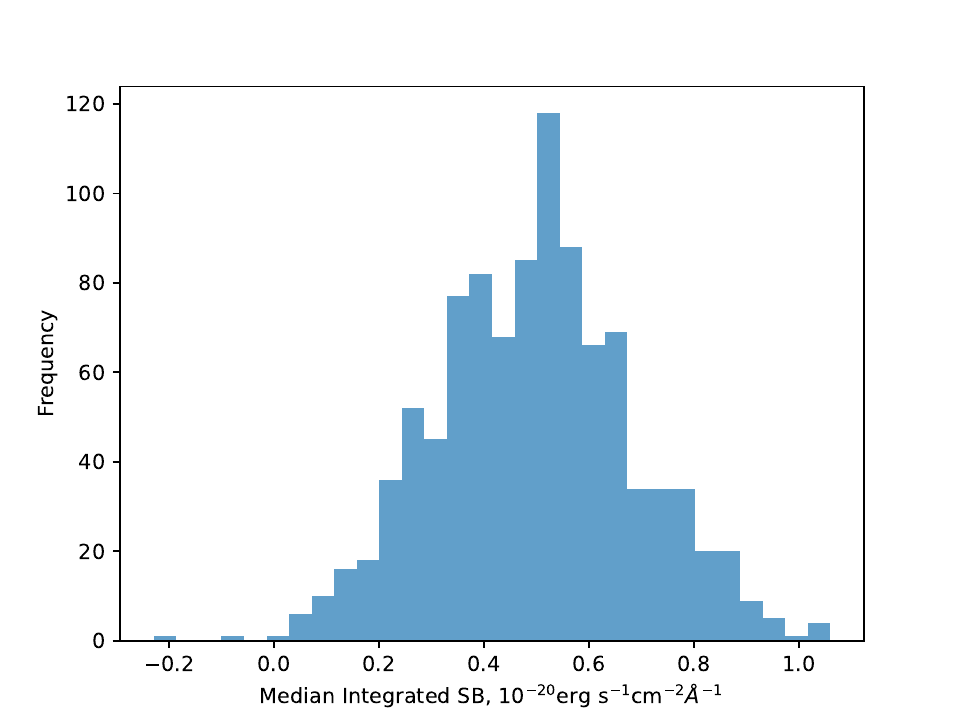}
    \caption{Jackknife of region 3 in Fig. \ref{fig:v1_regions_ro1001}}
    \label{fig:jackknife}
\end{figure*}

To inspect the noise level across the two fields in the background-subtracted data cubes after noise re-normalisation, we placed apertures, sized as the average of the box apertures in each field as shown in Fig. \ref{fig:v1_specs_bridges}, at random locations avoiding the LANs and the regions between the two pairs. The S/N distributions of pixels within these apertures in the two fields are shown in Fig. \ref{fig:snr_regions}. We find that the S/N distribution of the sky background is close to a Gaussian in both fields, and thus we do not expect many spurious positive pixels. We then performed a jackknife resampling on pixels within region 3 in Fig. \ref{fig:v1_specs_bridges} as shown in Fig. \ref{fig:jackknife}. The median integrated SB is positively skewed, suggesting the presence of diffuse emission in this region.

\end{appendix}

\end{document}